\documentclass[10pt,twocolumn,journal]{IEEEtran}
\usepackage{latexsym}
\usepackage{amsfonts}
\usepackage{amsbsy}
\usepackage{amsmath,amssymb}
\usepackage{amsthm}
\usepackage{times}
\usepackage{graphicx}
\usepackage{enumerate}
\usepackage[usenames]{color}
\usepackage[dvips]{pstcol}
\usepackage{epstopdf}
\usepackage{cite}
\usepackage{amsmath}
\usepackage{amssymb}
\usepackage{amsfonts}
\usepackage{graphicx}
\usepackage{epsfig}
\usepackage{psfrag}
\usepackage{xcolor}
\usepackage{amsfonts, bm}
\usepackage{epstopdf}
\usepackage{cite}
\usepackage{color}
\usepackage{xcolor}
\usepackage{verbatim}
\usepackage{subfig}
\usepackage{algorithm}
\usepackage{cases}
\usepackage{array}
\usepackage{algorithmic}
\usepackage{booktabs}
\newtheorem{remark}{Remark}

\usepackage{multirow}
\usepackage{stfloats}

\input epsf
\begin{document}
\title{Deep-Unfolding Beamforming for Intelligent Reflecting Surface assisted Full-Duplex Systems}


\vspace{-0.1em}
\author{Yanzhen Liu, Qiyu Hu, Yunlong Cai, \IEEEmembership{Senior Member,~IEEE,} Guanding Yu, \IEEEmembership{Senior Member,~IEEE,} and Geoffrey Ye Li, \IEEEmembership{Fellow,~IEEE}
\thanks{
Y. Liu, Q. Hu, Y. Cai,  and G. Yu  are with the College of Information Science and Electronic Engineering, Zhejiang University, China (e-mail:  yanzliu@zju.edu.cn; qiyhu@zju.edu.cn;  ylcai@zju.edu.cn; yuguanding@zju.edu.cn). G. Y. Li is with the Department of Electrical and Electronic Engineering, Imperial College London, London, UK (e-mail: geoffrey.li@imperial.ac.uk).

}
}

\maketitle
\vspace{-2.2em}
\begin{abstract}
In this paper, we investigate an intelligent reflecting surface (IRS) assisted multi-user multiple-input multiple-output (MIMO) full-duplex (FD) system. 
We jointly optimize the active beamforming matrices at the access point (AP) and uplink users, and the passive beamforming matrix at the IRS to maximize the weighted sum-rate of the system. Since it is practically difficult to acquire the channel state information (CSI) for IRS-related links due to its passive operation and large number of elements, we conceive a mixed-timescale beamforming scheme.
Specifically, the high-dimensional passive beamforming matrix at the IRS is updated based on the channel statistics while the active beamforming matrices are optimized relied on the low-dimensional real-time effective CSI at each time slot.
We propose an efficient stochastic successive convex approximation (SSCA)-based algorithm for jointly designing the active and passive beamforming matrices. Moreover, due to the high computational complexity caused by the matrix inversion computation in the SSCA-based optimization algorithm, we further develop a deep-unfolding neural network (NN) to address this issue. The proposed deep-unfolding NN maintains the structure of the SSCA-based algorithm but introduces a novel non-linear activation function and some learnable parameters induced by the first-order Taylor expansion to approximate the matrix inversion. In addition, we develop a black-box NN as a benchmark. Simulation results show that the proposed mixed-timescale algorithm outperforms the existing single-timescale algorithm and the proposed deep-unfolding NN approaches the performance of the SSCA-based algorithm with much reduced computational complexity when deployed online. 
\end{abstract}
\vspace{-0.1em}

\IEEEpeerreviewmaketitle

\vspace{-1.2em}
\section{Introduction}
\label{sec:intro}
Intelligent reflecting surface (IRS)~\cite{IRSintro1,IRSintro2,IRSintro3} can enhance the network throughput and has received great attention recently. Specifically, IRS, a planar meta-surface, is composed of a large number of tunable reflecting elements, each of which is able to induce the incident signal via changing its phase and/or amplitude. Based on the channel state information (CSI) and/or the channel statistics, the central controller of the IRS can collaboratively adjust its reflection coefficients such that the desired and interfering signals can be enhanced and suppressed, respectively, thus substantially improving the wireless  system performance. Moreover, compared to conventional techniques, such as active relaying/beamforming, IRS does not require any transmit/receive radio frequency (RF) chains. Hence, it is also more energy  and hardware efficient.

Full-duplex (FD), another powerful wireless technique, can significantly improve the spectral efficiency (SE) \cite{FDintro1,FDintro2,FDintro3,FDintro4}. Compared with the conventional half-duplex (HD) scheme, it fully utilizes the spectrum by enabling signal transmission and reception over the  same frequency at the same time, which can double the SE theoretically. However, the self-interference (SI) caused by the simultaneous downlink (DL) and uplink (UL) transmission is a challenging issue in FD systems. Fortunately, there have been key advances to address this issue~\cite{FDintro4,FDintro5,FDintro6}, such as passive
suppression, analog and digital cancellations, etc.. Therefore, the FD technique has many applications in wireless communications, such as bidirectional communications \cite{FDbidirectional}, relays \cite{FDrelay1,FDrelay2}, and multi-user systems \cite{FDmultiuser1,FDmultiuser2}.

\vspace{-1.0em}
\subsection{Prior works}
Lately, there have been a number of applications of IRS in various wireless communication scenarios, such as the IRS-aided secure communication~\cite{IRSsecure1,IRSsecure2,IRSsecure3}, simultaneous wireless information and power transfer~\cite{IRSswipt1,IRSswipt2}, millimeter wave communication~\cite{IRSmmWave1,IRSmmWave2}, and mobile edge computing~\cite{IRSMEC1}. More recently, several pieces of works  on IRS-aided FD systems  have been proposed~\cite{IRSFD1,IRSFD2,IRSFD3,IRSFD4,IRSFD5,IRSFD6}. In \cite{IRSFD1}, an IRS is used to enhance FD two-way communication systems, where the source precoders and the IRS passive beamforming matrix are jointly optimized to maximize the system capacity based on the Arimoto-Blahut algorithm. Under the same scenario, an algorithm with faster convergence speed has been proposed in \cite{IRSFD2}. Moreover, a novel hybrid communication network that utilizes both an FD decode-and-forward relay and an IRS to enhance data transmission rate has been investigated in \cite{IRSFD3}. 
In addition, the passive beamforming and deployment design have been investigated in \cite{IRSFD4} in an IRS-aided cellular FD system. To ensure user fairness, the minimum weighted rate is maximized in \cite{IRSFD5} in an IRS-aided multi-user cellular  FD system. Furthermore, the resource allocation problem for an IRS-assisted FD cognitive radio system has been studied in \cite{IRSFD6}.

It is worth mentioning that in the above studies, the beamforming matrices are designed based on the instantaneous CSI, which will incur high computational complexity and large signaling overhead in practice. Recently, more practical schemes have been developed by exploiting the channel statistics~\cite{IRSCDI1,IRSCDI2,IRSCDI3}. The angle domain framework in \cite{IRSCDI1} designs the beamforming matrices at the access point (AP) and IRS based on the derived effective angles, which approaches the performance with full CSI. By utilizing historical channel samples, the authors of \cite{IRSCDI2} proposed two stochastic optimization algorithms to configure the IRS phase shifter. In \cite{IRSCDI3}, a two-timescale protocol has been exploited to design the passive beamforming matrix based on the channel statistics and the active beamforming matrices  based on the effective CSI.

Although the aforementioned algorithms that utilize channel statistics  can significantly reduce the CSI overhead, they are still with high computational complexity since complex manipulations, such as matrix inversion, are involved in each iteration. In order to tackle this problem, machine learning based techniques, such as deep neural network (DNN), have been employed for beamforming design in IRS-aided systems~\cite{blackbox1,blackbox2,blackbox3}. DNN  only consists of linear operation and simple non-linear activation function, which can potentially meet the real-time requirement~\cite{fastbeamforming}. However, the black-box NNs generally have poor interpretability and require a large number of training samples. To this end, deep-unfolding NN~\cite{unfoldingintro1} unfolds some iterative optimization algorithms into layer-wise structures and learns the key parameters. The deep-unfolding NNs take advantages of both the model-driven optimization  algorithms and the data-driven learning-based algorithms. They are more interpretable and efficient than the black-box NNs and can achieve comparable performance with the conventional optimization algorithms with dramatically reduced computational complexity. Hence, the deep-unfolding has attracted great research interests and has a wide range of applications in communications, such as signal detection~\cite{unfolding1,unfolding2,unfolding3}, resource allocation~\cite{unfolding4,unfolding5}, and precoding~\cite{unfolding6,unfolding7,unfolding8,unfolding9}. 

\vspace{-1.0em}
\subsection{Main Contributions}
Inspired by the above works, we investigate a multi-user MIMO IRS-assisted FD system in this paper, which consists of an AP, an IRS, multiple UL users and DL users as shown in Fig. \ref{fig:structure}. The AP operates in an FD mode and the users operate in an HD mode. We jointly optimize the active beamforming matrices at the AP and UL users and the passive beamforming matrix at the IRS to maximize the weighted sum-rate of the system. 
Since it is practically difficult to acquire the CSI for IRS-related links due to its passive operation and large number of elements, we conceive a mixed-timescale scheme. 
Specifically, the high-dimensional passive beamforming matrix at the IRS is updated based on the channel statistics while the active beamforming matrices are optimized relied on the low-dimensional real-time effective CSI at each time slot. The proposed scheme avoids estimating the high dimensional IRS-related channels in each time slot and saves the heavy overhead required by the conventional single-timescale algorithm, thus alleviating the performance degradation caused by CSI delay. 

However, the mixed-timescale brings new challenges to algorithm design since the objective function turns stochastic and the long-term and short-term variables are highly coupled. To address these issues, we develop an efficient stochastic successive convex approximation (SSCA)-based optimization algorithm. More precisely, for the short-term active beamforming design, we equivalently convert the original problem into a more tractable form and propose a block coordinate descent (BCD)-type algorithm. Then, with optimized short-term active beamforming matrices, the long-term passive beamforming matrix is designed based on SSCA~\cite{cssca}. Specifically, we construct a convex surrogate function based on the collected full CSI samples\:\!\footnote{In this work, channel statistics refer to the moments or distribution of the
channel fading realizations. By observing the collected full channel samples (possibly outdated), the proposed SSCA-based design algorithm can automatically learn the channel statistics (in an implicit way) and converge to
a stationary point of the stochastic optimization problem considered in our
design.} to approximate the objective function and iteratively optimize the IRS passive beamforming matrix until convergence. The proposed algorithm can be guaranteed to converge to a stationary point of the original problem.

Furthermore, we design a novel deep-unfolding NN that jointly unfolds the proposed SSCA-based optimization algorithm into a layer-wise structure. The proposed deep-unfolding NN consists of a long-term passive beamforming network (LPBN) and a short-term active beamforming network (SABN). In the forward propagation stage, the collected full channel samples are first fed into the LPBN and it outputs the low-dimensional effective CSI. Note that we directly set the IRS passive beamforming matrix as the learnable parameter of the LPBN. Then, the effective CSI passes through the SABN and it outputs the active beamforming matrices. The SABN maintains the structure of the proposed BCD-type active beamforming algorithm but employs a novel non-linear activation function and some learnable parameters induced by the first-order Taylor expansion to approximate the matrix inversion, which significantly reduces the computational complexity. In the back propagation stage, the learnable parameters of the deep-unfolding NN are updated based on the stochastic gradient descent (SGD) method. The main contributions of this paper are summarized as follows.

\begin{itemize}
	\item We study a multi-user MIMO IRS-assisted FD system, which has not been well investigated in the literature,  and propose a practical mixed-timescale beamforming scheme to reduce the CSI overhead and mitigate the CSI mismatch caused by delay.
    
    \item To maximize the weighted average sum-rate in the IRS-assisted FD system, we propose an efficent mixed-timescale joint active and passive beamforming algorithm  based on the framework of SSCA, which can guarantee convergence.
    
    \item To further reduce the computational complexity, we develop a novel deep-unfolding NN that unfolds the proposed mixed-timescale SSCA-based algorithm into a layer-wise structure. The deep-unfolding NN exploits a novel non-linear activation function and some learnable parameters induced by the first-order Taylor expansion  to approximate the matrix inversion. 
    
    \item Simulation results show that the proposed mixed-timescale beamforming algorithm outperforms the single-timescale counterpart in the presence of CSI delay, and the proposed deep-unfolding NN approaches the performance of the SSCA-based algorithm with much reduced computational complexity when deployed online.   	   
\end{itemize}
\vspace{-1.0em}
\subsection{Organizations and Notations}
The paper is organized as follows. Section II describes the system model and formulates the investigated problem. The proposed SSCA-based  mixed-timescale beamforming optimization algorithm is presented in Section III. Then, Section IV introduces the proposed deep-unfolding NN based algorithm. Section V presents the simulation results and Section VI concludes this paper. 

Scalars, vectors and matrices are denoted by lower case, boldface lower case and boldface upper case letters, respectively. $\mathbf{I}$ represents an identity matrix and $\mathbf{0}$ denotes an all-zero  matrix.
For a  matrix $\mathbf{A}$, ${{\mathbf{A}}^T}$, $\textrm{conj}(\mathbf{A})$, ${{\mathbf{A}}^H}$, and $\|\mathbf{A}\|$ denote its transpose, conjugate, conjugate transpose, and Frobenius norm, respectively. For a square matrix $\mathbf{A}$, $\textrm{Tr} \{\mathbf{A}\}$ and $\mathbf{A}^{-1}$ denotes its trace and inverse, respectively, while ${\mathbf{A}} \succeq {\mathbf{0}}~({\mathbf{A}} \preceq {\mathbf{0}})$ means that $\mathbf{A}$ is  positive (negative) semi-definite.
For a vector $\mathbf{a}$, $\|\mathbf{a}\|$ represents its Euclidean norm.
$\Re e\{\cdot\}$ ($\Im m\{\cdot\}$) denotes the real  (imaginary) part of a variable.
$|  \cdot  |$ denotes  the absolute value of a complex scalar.
${\mathbb{C}^{m \times n}}\;({\mathbb{R}^{m \times n}})$ denotes the space of ${m \times n}$ complex (real) matrices and $\angle$ represents the phase of complex vectors/matrices. $\text{diag}(\cdot)$ extracts the diagonal elements of a square matrix and $\text{Diag}(\cdot)$ constructs a diagonal matrix based on the input vector. The key notations used in this paper is summarized in Table I.
\vspace{-0.2em}

\begin{table}[htbp]
	\centering  
	\caption{List of notations.}  
	\label{table1}  
	\begin{tabular}{|c|c|}  
		\hline  
		Symbol&Representation\\  
		\hline
		$K$ ($k$)&Number of UL users (index for UL users)\\
		\hline
		$L$ ($l$)&Number of DL users (index for DL users) \\
		\hline
		$N_t$ ($N_r$)&Number of transmit (receive) antennas at the AP\\
		\hline
		$M_\text{U}$ ($M_\text{D}$)  &Number of antennas at the UL (DL) users \\
		\hline
		$D_{\text{U}}$ ($D_{\text{D}}$)&Number of data flows at the UL (DL) users\\
		\hline
		$T$ &Number of reflecting elements at the IRS \\  
		\hline
		$\mathbf{\Phi}$&Passive beamforming matrix at the IRS (long-term)\\
		\hline
		$\mathbf{P}$ &Active beamforming matrix at UL user (short-term) \\		

		\hline
		$\mathbf{F}$&Active beamforming matrix for DL user (short-term)\\
		\hline
		$\mathcal{H}$ ($\mathcal{H}_{ef}$) &Set of full (effective) channels \\
		\hline
		$R_{\text{U}}$ ($R_{\text{D}}$)&Achievable rate of UL (DL) user \\
		\hline
		$\alpha$ ($\beta$) &Weight of UL (DL) user \\
		\hline
		$\boldsymbol{\phi}$&Diagonal elements of $\mathbf{\Phi}$ \\
		\hline
		$\boldsymbol{\theta}$ &Phase of $\boldsymbol{\phi}$ \\
		\hline
	\end{tabular}
\end{table}

\vspace{-1.6em}
\section{System Model and Problem Formulation}
In this section, we first introduce the IRS-assisted FD system and then mathematically formulate the optimization problem.
\vspace{-1.6em}
\subsection{System Model}
\begin{figure}[!t]
\centering
\scalebox{0.50}{\includegraphics{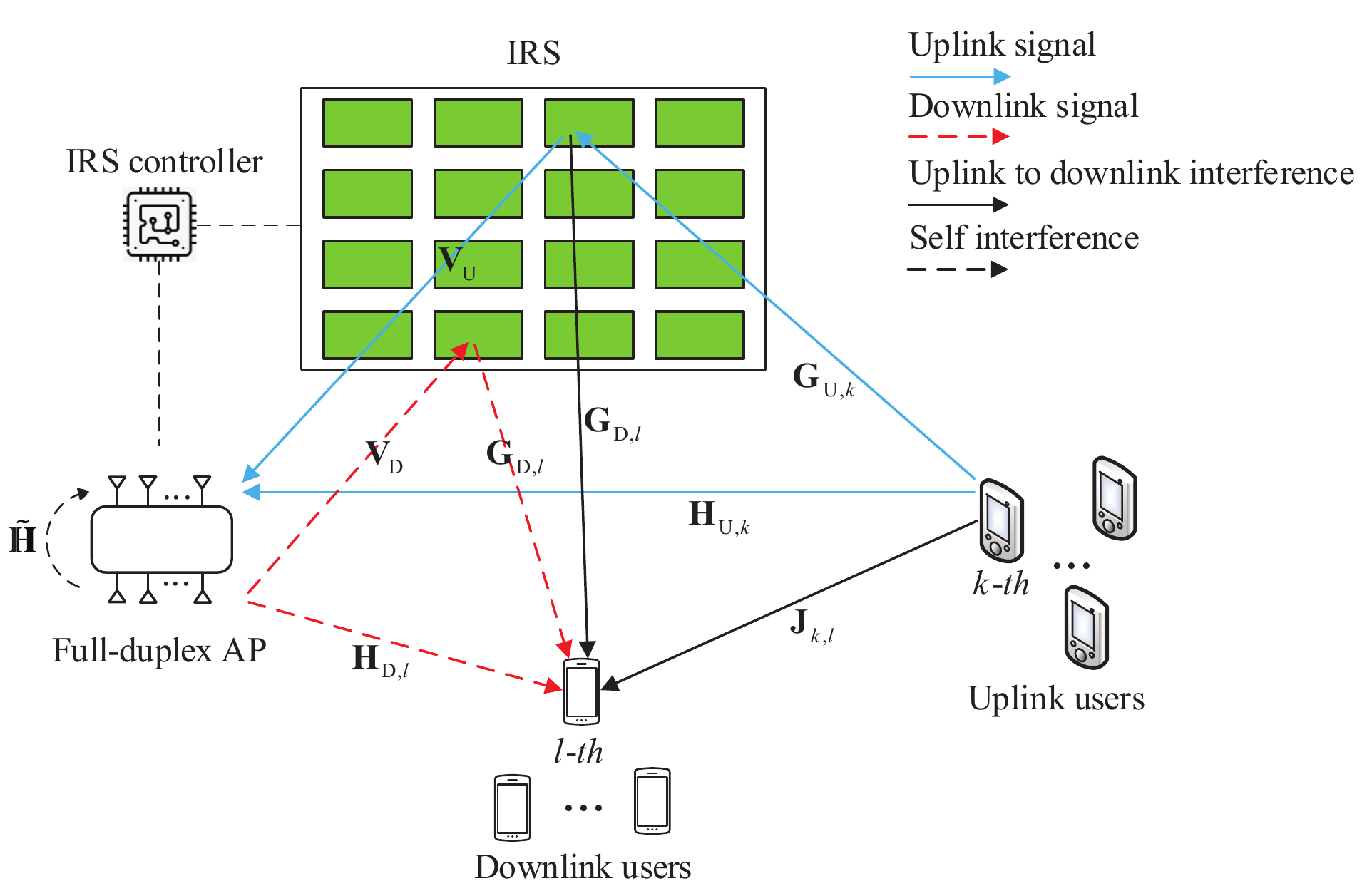}}
\caption{ IRS-assisted full-duplex system.}\label{fig:structure}
\end{figure}

As depicted in Fig.~\ref{fig:structure}, we consider an IRS-assisted FD system, which consists of an AP, an IRS, $K$ UL users, and $L$ DL users. The AP operates in an FD mode and it is equipped with $N_t$ transmit antennas and $N_r$ receive antennas. The $K$ UL users and the $L$ DL users operate in an HD mode and they are equipped with $M_{\text{U}, k}$ and $M_{\text{D}, l}$ antennas, respectively, where $k\in\mathcal{K}\triangleq\{1, \ldots, K\}$ and $l\in\mathcal{L}\triangleq\{1,\ldots, L\}$ denote the user indexes. The IRS is equipped with $T$ reflecting elements.

Assuming that the IRS is equipped near the users and far away from the AP, the signals through the AP-IRS-AP link can be neglected due to the high path loss~\cite{IRSFD4}. Then,
the received data vector at the AP $\mathbf{y}_\text{U}\in\mathbb{C}^{N_r\times 1}$ is given by
\begin{equation}
\mathbf{y}_\text{U}=\sum^{K}_{k=1}\mathbf{H}_{\text{U},k}\mathbf{P}_{k}\mathbf{b}_{k}+\sum^{K}_{k=1}\mathbf{V}_{\text{U}}\mathbf{\Phi}\mathbf{G}_{\text{U},k}\mathbf{P}_{k}\mathbf{b}_{k}+\sum^{L}_{l=1}\mathbf{\tilde{H}}\mathbf{F}_{l}\mathbf{s}_{l}+\mathbf{n}_\text{U},
\end{equation}
where $\mathbf{b}_{k}\in\mathbb{C}^{D_{\text{U},k}\times 1}$ ($D_{\text{U},k}\leq M_{\text{U},k}$) denotes the transmit data vector of the $k$-th UL user, $\mathbf{P}_{k}\in\mathbb{C}^{M_{\text{U},k}\times D_{\text{U},k}}$ is the beamforming matrix of the $k$-th UL user, $\mathbf{H}_{\text{U},k}\in\mathbb{C}^{N_r\times M_{\text{U},k}}$ denotes the channel matrix between the $k$-th UL user and the AP.
$\mathbf{\Phi}\in\mathbb{C}^{T\times T}$ denotes the diagonal passive beamforming matrix at the IRS due to no signal coupling/joint processing over its passive reflecting elements, $\mathbf{G}_{\text{U},k}\in\mathbb{C}^{T\times M_{\text{U},k}}$ denotes the channel matrix between the $k$-th UL user and the IRS, and $\mathbf{V}_{\text{U}}\in\mathbb{C}^{N_r\times T}$ denotes the channel matrix between the IRS and the AP. Note that
$\mathbf{s}_{l}\in\mathbb{C}^{D_{\text{D}, l}\times 1}$ ($D_{\text{D}, l}\leq M_{\text{D}, l}$) denotes the data vector for the $l$-th DL user,
$\mathbf{F}_{l}\in\mathbb{C}^{N_t\times D_{\text{D},l}}$ denotes the beamforming matrix at the AP for serving the $l$-th DL user, and $\mathbf{\tilde{H}}\in\mathbb{C}^{N_r\times N_t}$ denotes the residual SI channel matrix at the AP\:\!\footnote{Since the CSI of the SI link can be obtained at the AP, based on certain interference cancellation techniques~\cite{FDintro3,FDintro6}, we assume that the SI at the AP can be greatly eliminated.}.
$\mathbf{n}_\text{U}\in\mathbb{C}^{N_r\times 1}$ denotes the complex circular Gaussian noise vector at the AP with zero mean and variance $\sigma^2_\text{U}$.

The received data vector at the $l$-th DL user $\mathbf{y}_{\text{D},l}\in\mathbb{C}^{M_{\text{D},l}\times 1}$ is given by 
 \begin{equation}
 	\begin{split}
 \mathbf{y}_{\text{D},l}=&\sum^{L}_{l=1}\mathbf{H}_{\text{D},l}\mathbf{F}_{l}\mathbf{s}_{l}+\sum^{L}_{l=1}\mathbf{G}_{\text{D},l}\mathbf{\Phi}\mathbf{V}_\text{D}\mathbf{F}_{l}\mathbf{s}_{l}\\
 &+\underbrace{\sum^{K}_{k=1}\mathbf{J}_{k,l}\mathbf{P}_{k}\mathbf{b}_{k}+\sum^{K}_{k=1}\mathbf{G}_{\text{D},l}\mathbf{\Phi}\mathbf{G}_{\text{U},k}\mathbf{P}_{k}\mathbf{b}_{k}}_{interference\,\, from\,\, the\,\, UL\,\, users}+\mathbf{n}_{\text{D},l},
  	\end{split}
 \end{equation}
 where  $\mathbf{H}_{\text{D},l}\in\mathbb{C}^{M_{\text{D},l}\times N_t}$ is the channel matrix between the AP and the $l$-th DL user, $\mathbf{V}_{\text{D}}\in\mathbb{C}^{T\times N_t}$ denotes the channel matrix between the AP and the IRS, $\mathbf{G}_{\text{D},l}\in\mathbb{C}^{M_{\text{D},l}\times T}$ is the channel matrix between the IRS and the $l$-th DL user, $\mathbf{J}_{k,l}\in\mathbb{C}^{M_{\text{D},l}\times M_{\text{U},k}}$ denotes the channel matrix between the $k$-th UL user and the $l$-th DL user, and $\mathbf{n}_{\text{D},l}\in\mathbb{C}^{M_{\text{D},l}\times 1}$ denotes the complex circular Gaussian noise vector at the $l$-th DL user with zero mean and variance $\sigma^2_{\text{D},l}$.

The transmission rate for user $k$ in the uplink is given by \vspace{-0.4em}
\begin{equation}
	\vspace{-0.5em}
	\begin{split}
	&\mathcal{R}_{\text{U},k}\triangleq  \log \det \bigg( \mathbf{I}+\mathbf{\bar{H}}_{\text{U},k}\mathbf{P}_{k}\mathbf{P}^H_{k}\mathbf{\bar{H}}^H_{\text{U},k}\times\\ 
	&(\sum^{L}_{l=1}\mathbf{\tilde{H}}\mathbf{F}_{l}\mathbf{F}^H_{l}\mathbf{\tilde{H}}^H+\sum^{K}_{k^{'}\neq k} \bar{\mathbf{H}}_{\text{U},k^{'}}\mathbf{P}_{k^{'}}\mathbf{P}^H_{k^{'}}\bar{\mathbf{H}}^H_{\text{U},k^{'}} +\sigma^2_\text{U}\mathbf{I})^{-1} \bigg),
\end{split}
\end{equation}
where $\mathbf{\bar{H}}_{\text{U}, k}\triangleq \mathbf{H}_{\text{U},k}+\mathbf{V}_\text{U}\mathbf{\Phi}\mathbf{G}_{\text{U},k}$. 

The transmission rate for user $l$ in the downlink is given by \vspace{-0.4em}
\begin{equation}
	\vspace{-0.4em}
	\begin{split}
	&\mathcal{R}_{\text{D},l}\triangleq\log \det \bigg( \mathbf{I}+\mathbf{\bar{H}}_{\text{D},l}\mathbf{F}_{l}\mathbf{F}^H_{l}\mathbf{\bar{H}}^H_{\text{D},l}\times \\
	&(\sum^{K}_{k=1}\mathbf{\bar{J}}_{k,l}\mathbf{P}_{k}\mathbf{P}^H_{k}\mathbf{\bar{J}}^H_{k,l}+\sum^{L}_{l^{'}\neq l}\mathbf{\bar{H}}_{\text{D},l^{'}}\mathbf{F}_{l^{'}}\mathbf{F}^H_{l^{'}}\mathbf{\bar{H}}^H_{\text{D},l^{'}}+\sigma^2_{\text{D},l}\mathbf{I})^{-1} \bigg),
	\end{split}
\end{equation}
where $\mathbf{\bar{H}}_{\text{D},l}\triangleq \mathbf{H}_{\text{D},l}+\mathbf{G}_{\text{D},l}\mathbf{\Phi}\mathbf{V}_\text{D}$ and 
$\mathbf{\bar{J}}_{k,l}\triangleq \mathbf{J}_{k,l}+\mathbf{G}_{\text{D,l}}\mathbf{\Phi}\mathbf{G}_{\text{U},k}$.
\vspace{-0.2em}
\subsection{Mixed-Timescale Protocols}
\begin{figure*}[!t]
	\centering
	\includegraphics[width=0.76\linewidth,height=0.25\linewidth]{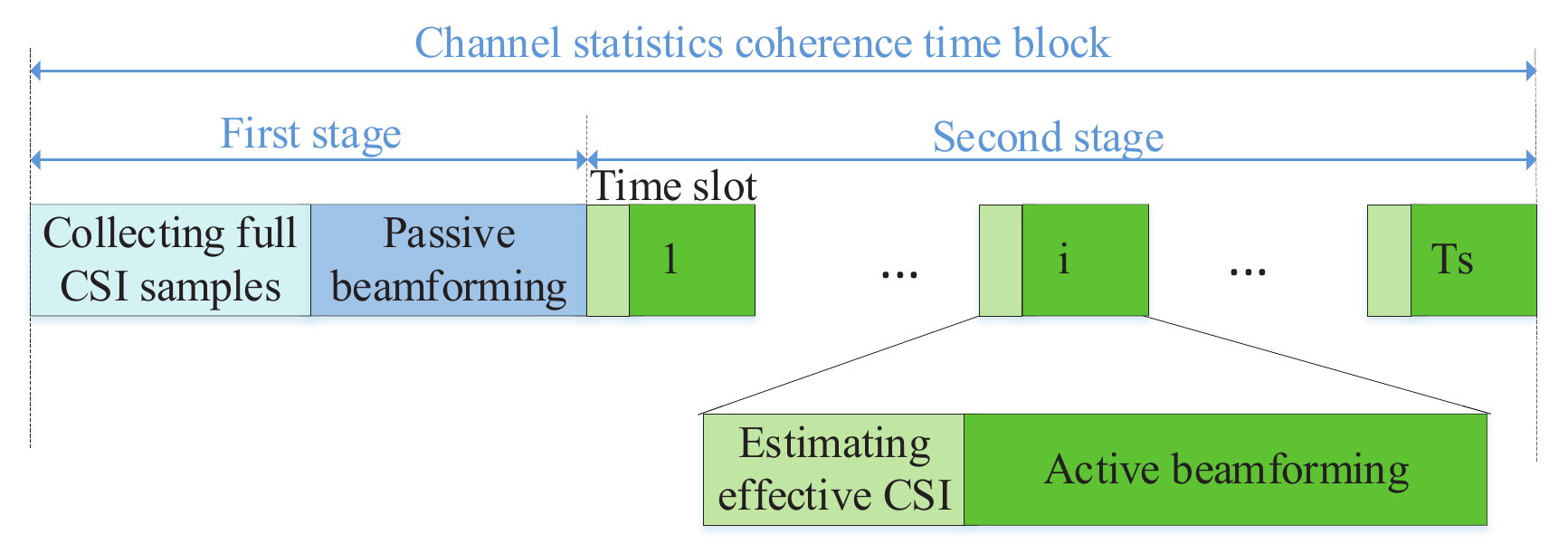}
	\caption{ Proposed mixed-timescale beamforming scheme.}\label{fig:timescale} \vspace{-0.6em}
\end{figure*}
In practice, the acquisition of the real-time IRS-related high-dimensional CSI matrix is very challenging due to the large number of reflecting elements and the passive architecture of the IRS while estimating the low-dimensional effective channels $\mathcal{H}_{ef} \triangleq \{\bar{\mathbf{H}}_{\text{U},k},\bar{\mathbf{H}}_{\text{D},l},\bar{\mathbf{J}}_{k,l},\tilde{\mathbf{H}}\}$ with given IRS passive beamforming matrix is much easier. Based on this observation, we propose a mixed-timescale transmission protocol. Specifically, we focus on a sufficient large time block during which the channel statistics are constant, as shown in Fig. \ref{fig:timescale}. In the first stage, the AP estimates a small amount of high-dimensional full CSI samples $\{\mathcal{H}(n)\}_{n=\{1,...,N_{s}\}}$  (possibly outdated) using some standard IRS-related channel estimation methods~\cite{IRS_CM1,IRS_CM2,IRS_CM3}, where $N_{s}$ denotes the number of collected full CSI samples, and $\mathcal{H} \triangleq \{\mathbf{H}_{\text{U},k},\mathbf{H}_{\text{D},l},\mathbf{G}_{\text{U},k},\mathbf{G}_{\text{D},l},\mathbf{V}_{\text{U}},\mathbf{V}_{\text{D}},\mathbf{J}_{k,l},\tilde{\mathbf{H}}\}$ denotes the set of all CSI matrices. Then, the AP designs the long-term passive beamforming matrix $\mathbf{\Phi}$ based on these collected full CSI samples and sends it to the IRS.

In the second stage, the long-term passive beamforming matrix at the IRS is fixed.  In each time slot (channel coherence time) $i$, the AP obtains the low-dimensional effective channels $\mathcal{H}_{ef}(i)$ via conventional MIMO channel estimation methods and designs the active short-term beamforming matrices $\mathbf{P}_{k}$ and $\mathbf{F}_{l}$ accordingly. Then, the beamforming matrices $\mathbf{P}_{k}$ are sent to the UL users. As we can see, the proposed mixed-timescale scheme avoids estimating the high-dimensional IRS-related CSI matrix in each time slot. By contrast, the conventional single-timescale algorithm requires a tremendous amount of full CSI samples in each coherence time block, which needs a huge overhead.   

\vspace{-0.5em}
\subsection{Problem Formulation}

In this work, we aim at jointly designing the short-term active beamforming matrices $\mathbf{P}_{k}$ and $\mathbf{F}_{l}$, and the long-term IRS passive beamforming matrix $\mathbf{\Phi}$ in order to maximize the average weighted  sum-rate over a coherence time block.
Hence, the problem can be formulated as follows \vspace{-1.8em}

 \begin{subequations} \label{eq:problem1}
 \begin{align}\vspace{-1.6em}
 (\mathcal{P}1): \quad \max_{\boldsymbol{\phi}} \mathbb{E}_{\mathcal{H}}\{\max_{\{\mathbf{P}_{k},\mathbf{F}_{l}\}} & \sum^{K}_{k=1}\alpha_{k} \mathcal{R}_{\text{U},k}+ \sum^{L}_{l=1}\beta_{l} \mathcal{R}_{\text{D},l}\} \label{objective}\\
 \mbox{s.t.}\quad
 &\|\mathbf{P}_{k}\|^2\leq P_{\text{U},k}, \, \forall k, \label{transmitpower}\\
 &\sum^{L}_{l=1}\|\mathbf{F}_{l}\|^2\leq P_{AP},\label{transmitpower2}\\
  & |\boldsymbol{\phi}(n)|=1, \forall  n, \label{constantmodulus}
 \end{align} 
 \end{subequations} 
where $\boldsymbol{\phi} \triangleq \textrm{diag}(\mathbf{\Phi})  \in\mathbb{C}^{T\times 1} $ denotes the long-term passive beamforming vector. $P_{\text{U},k}$ and $P_{AP}$ denote the limited transmit
power budgets of the UL users and the AP, respectively, and
\eqref{constantmodulus}  denotes the constant modulus constraint imposed on the elements of the IRS passive beamforming vector.

\vspace{-1.0em}
\section{Mixed-Timescale Beamforming Algorithm}
In this section, we propose the mixed-timescale beamforming algorithm for solving $\mathcal{P}1$. Firstly, by fixing the long-term passive beamforming matrix at the IRS, we optimize the short-term active beamforming matrices at the AP and UL users, where a BCD-type algorithm is proposed to tackle this problem. Then, we develop an efficient SSCA-based algorithm for designing the long-term passive beamforming matrix at the IRS.
\vspace{-1.0em}
\subsection{Short-Term Active Beamforming Design}
With fixed long-term IRS passive beamforming matrix, the optimization problem of the short-term active beamforming design is given by
\begin{subequations} \label{eq:short_problem}
	\begin{align}
		(\mathcal{P}2): \quad \max_{\{\mathbf{P}_{k},\mathbf{F}_{l}\} }\quad & \sum^{K}_{k=1}\alpha_{k} \mathcal{R}_{\text{U},k}+ \sum^{L}_{l=1}\beta_{l} \mathcal{R}_{\text{D},l} \label{shortobjective}\\
		\mbox{s.t.}\quad
		&\|\mathbf{P}_{k}\|^2\leq P_{\text{U},k}, \, \forall k, \label{shorttransmitpower}\\
		&\sum^{L}_{l=1}\|\mathbf{F}_{l}\|^2\leq P_{AP}.\label{shorttransmitpower2}
	\end{align}
\end{subequations}
The objective function of $\mathcal{P}2$ is difficult to handle due to the highly nonlinear objective function and coupled optimization variables. Hence, we first transform $\mathcal{P}2$ into an equivalent but more tractable form via the celebrated WMMSE method~\cite{WMMSE}. Specifically, we introduce auxiliary variables $\mathbf{W}_{\text{U},k} \in \mathbb{C}^{D_{\text{U},k}\times D_{\text{U},k}}$,$\mathbf{W}_{\text{D},l} \in \mathbb{C}^{D_{\text{D},l}\times D_{\text{D},l}}$,$\mathbf{U}_{\text{U},k} \in \mathbb{C}^{N_r\times D_{\text{U},k}}$, and $\mathbf{U}_{\text{D},l} \in \mathbb{C}^{M_{\text{D},l}\times D_{\text{D},l}}$, and the converted problem can be expressed as \vspace{-0.2em}
\begin{subequations} \vspace{-0.2em}
	\begin{align}
		(\mathcal{P}3): \min_{\Omega}\quad &\sum_{k=1}^{K} \alpha_k\left(\text{Tr}\left(\mathbf{W}_{\text{U},k}\mathbf{E}_{\text{U},k}\right)-\log \det\left(\mathbf{W}_{\text{U},k}\right)\right) \nonumber\\
		+\sum_{l=1}^{L} &\beta_l\left(\text{Tr}\left(\mathbf{W}_{\text{D},l}\mathbf{E}_{\text{D},l}\right)-\log \det\left(\mathbf{W}_{\text{D},l}\right)\right) \\
		\mbox{s.t.}\quad& \eqref{transmitpower}, \eqref{transmitpower2}, 
	\end{align} 
\end{subequations}
where $\Omega \triangleq \{\mathbf{P}_k,\mathbf{F}_l,\mathbf{U}_{\text{U},k},\mathbf{U}_{\text{D},l},\mathbf{W}_{\text{U},k},\mathbf{W}_{\text{D},l}\}$ is the set of optimization variables, and \vspace{-0.3em}
\begin{equation} \vspace{-0.3em}
	\begin{split}
	&\mathbf{E}_{\text{U},k} \triangleq (\mathbf{U}_{\text{U},k}^{\rm H}\bar{\mathbf{H}}_{\text{U},k}\mathbf{P}_k-\mathbf{I})(\mathbf{U}_{\text{U},k}^{\rm H}\bar{\mathbf{H}}_{\text{U},k}\mathbf{P}_k-\mathbf{I})^{\rm H} + \mathbf{U}_{\text{U},k}^{\rm H}\times\\
	&\left(\sum_{k^{'}\neq k}^{K}\bar{\mathbf{H}}_{\text{U},k^{'}}\mathbf{P}_{k^{'}}\mathbf{P}_{k^{'}}^{\rm H}\bar{\mathbf{H}}_{\text{U},k^{'}}^{\rm H}+\sum_{l=1}^{L}\tilde{\mathbf{H}}\mathbf{F}_l\mathbf{F}_l^{\rm H}\tilde{\mathbf{H}}^{\rm H}+\sigma_{\text{U}}^2\mathbf{I}\right)\mathbf{U}_{\text{U},k},
    \end{split} 
\end{equation}
\begin{equation}
	\begin{split}
	&\mathbf{E}_{\text{D},l} \triangleq (\mathbf{U}_{\text{D},l}^{\rm H}\bar{\mathbf{H}}_{\text{D},l}\mathbf{F}_l-\mathbf{I})(\mathbf{U}_{\text{D},l}^{\rm H}\bar{\mathbf{H}}_{\text{D},l}\mathbf{F}_l-\mathbf{I})^{\rm H} + \mathbf{U}_{\text{D},l}^{\rm H}\times \\
	&\left(\sum_{l^{'}\neq l}^{L}\bar{\mathbf{H}}_{\text{D},l}\mathbf{F}_{l^{'}}\mathbf{F}_{l^{'}}^{\rm H}\bar{\mathbf{H}}_{\text{D},l}^{\rm H}+\sum_{k=1}^{K}\bar{\mathbf{J}}_{k,l}\mathbf{P}_k\mathbf{P}_k^{\rm H}\bar{\mathbf{J}}_{k,l}^{\rm H}+\sigma_{\text{D},l}^2\mathbf{I}\right)\mathbf{U}_{\text{D},l}.
\end{split}
\end{equation}

$\mathcal{P}2$ and $\mathcal{P}3$ are equivalent since they share the same global optimal solution \cite{WMMSE}. Then, we develop a BCD-type algorithm for solving $\mathcal{P}3$. Specifically, the set of optimization variables $\Omega$ is divided into four blocks, i.e. $\{\mathbf{U}_{\text{U},k},\mathbf{U}_{\text{D},l}\}$, $\{\mathbf{W}_{\text{U},k},\mathbf{W}_{\text{D},l}\}$, $\{\mathbf{P}_k\}$, and $\{\mathbf{F}_l\}$. Each block are optimized in turn with the other blocks of variables fixed. The proposed BCD-type algorithm for optimizing the short-term active beamforming matrices is summarized in Algorithm \ref{BCDtype}. The details on solving the subproblems w.r.t. each block of variables are given in Appendix A.

\begin{algorithm}[t]\caption{Proposed BCD-type short-term active beamforming design algorithm.} \label{BCDtype}
	\begin{algorithmic}[1]
		\footnotesize
		\begin{small}
			\STATE Initialize the beamforming matrices $\mathbf{P}_k$ and $\mathbf{F}_l$ with feasible values. Set the maximum iteration number $I_{max}$ and the threshold value $\delta$.
			\REPEAT
			\STATE  Update $\mathbf{U}_{{\rm U},k} = \mathbf{A}_{\text{U},k}^{-1} \bar{\mathbf{H}}_{\text{U},k}\mathbf{P}_k$ and $\mathbf{U}_{{\rm D},l} = \mathbf{A}_{\text{D},l}^{-1} \bar{\mathbf{H}}_{\text{D},l}\mathbf{F}_l$, where $\mathbf{A}_{\text{U},k}$ and $\mathbf{A}_{\text{D},l}$ are defined in~\eqref{AU} and~\eqref{AD}, respectively.
			\STATE  Update $\mathbf{W}_{\text{U},k} = \mathbf{E}_{\text{U},k}^{-1}$ and $\mathbf{W}_{\text{D},l} = \mathbf{E}_{\text{D},l}^{-1}$.
			\STATE Update $	\mathbf{P}_k = \alpha_k(\mathbf{A}_{\text{P},k}+\lambda_k\mathbf{I})^{-1} \bar{\mathbf{H}}_{\text{U},k}^{\rm H}\mathbf{U}_{\text{U},k}\mathbf{W}_{\text{U},k}$, where $\mathbf{A}_{\text{P},k}$ is defined in~\eqref{APk} and $\lambda_k$ is the Lagrange multiplier.
			\STATE Update $\mathbf{F}_l = \beta_l(\mathbf{A}_{\text{F}}+\mu \mathbf{I})^{-1}\bar{\mathbf{H}}_{\text{D},l}^{\rm H}\mathbf{U}_{\text{D},l}\mathbf{W}_{\text{D},l}$, where $\mathbf{A}_{\text{F}}$ is defined in~\eqref{AF} and $\mu$ is the Lagrange multiplier.
			\UNTIL the maximum iteration number is reached or the difference between the successive objective function value is less than $\delta$.
		\end{small}
	\end{algorithmic}
\end{algorithm}


\vspace{-0.4em}
\subsection{Long-term IRS Passive Beamforming Design}
In this subsection, we introduce the proposed long-term IRS passive beamforming design algorithm. With the optimized short-term variables, the stochastic optimization problem w.r.t. the passive beamforming vector can be formulated as 
\begin{equation}
	(\mathcal{P}4)\,\,\min_{\boldsymbol{\theta}} \quad f (\bm{\theta}, \{\mathbf{P}_k^*,\mathbf{F}_{l}^*\} ) = \mathbb{E}_{\mathcal{H}}\{g(\bm{\theta},\{\mathbf{P}_k^*,\mathbf{F}_{l}^*\};\mathcal{H})\}, \label{long-term-object}  \vspace{-1.5mm}
\end{equation}
where $\boldsymbol{\theta} \triangleq \angle{\boldsymbol{\phi}}$, $\{\mathbf{P}_k^*,\mathbf{F}_l^*\}$ is the optimal solution of the proposed short-term active beamforming algorithm and $g(\bm{\theta},\{\mathbf{P}_k^*,\mathbf{F}_{l}^*\};\mathcal{H})$ denotes the sum-rate given by \vspace{-0.2em}
\begin{equation}  \vspace{-0.2em}
	\begin{split}
	g(\bm{\theta},\{\mathbf{P}_k^*,\mathbf{F}_{l}^*\};\mathcal{H}) \triangleq &\sum^{K}_{k=1}\alpha_{k} \mathcal{R}_{\text{U},k}(\bm{\theta},\{\mathbf{P}_k^*,\mathbf{F}_{l}^*\};\mathcal{H})\\
	&+\sum^{L}_{l=1}\beta_{l}\mathcal{R}_{\text{D},l}(\bm{\theta},\{\mathbf{P}_k^*,\mathbf{F}_{l}^*\};\mathcal{H}).
	\end{split} \label{gfunction}
\end{equation}

Note that $\mathcal{P}4$ is hard to solve directly since the objective function is highly non-convex and it is difficult to obtain a closed-form expression via computing the expectation over $g(\bm{\theta},\{\mathbf{P}_k^*,\mathbf{F}_{l}^*\};\mathcal{H})$. Hence, by leveraging the stochastic optimization framework in~\cite{cssca}, we  approximate \eqref{long-term-object} by using a quadratic surrogate function. Specifically, at the $t$-th iteration of the proposed algorithm, $B$ channel samples, where $B$ is the batch size, denoted as $\{\mathcal{H}^t(m)\}_{m=\{1,...,B\}}$ are randomly selected from the collection of high-dimensional CSI samples $\{\mathcal{H}(n)
\}_{n=\{1,...,N_s\}}$ and the surrogate function is updated as \vspace{-0.2em}
\begin{equation} \vspace{-0.2em}
	\begin{split}
		\bar{f}^t(\boldsymbol{\theta}) =  (\mathbf{f}^t)^T(\boldsymbol{\theta}-\boldsymbol{\theta}^t)+\varpi\|\boldsymbol{\theta}-\boldsymbol{\theta}^t\|^2, \label{surrogate function}
	\end{split}
\end{equation}
where $\boldsymbol{\theta}^t$ denotes the current value of $\boldsymbol{\theta}$ and $\varpi>0$ is a constant. Note that $\mathbf{f}^t$ denotes the approximation of the partial derivatives $\frac{\partial f}{\partial \boldsymbol{\theta}}$, which is updated as \vspace{-0.2em}
\begin{equation} \vspace{-0.2em}
	\mathbf{f}^t = (1-\varrho^{t})\mathbf{f}^{t-1}+\varrho^t\sum_{m=1}^{B}\frac{\partial g(\boldsymbol{\theta},\{\mathbf{P}_k^*,\mathbf{F}_{l}^*\};\mathcal{H}^t(m)) }{\partial \boldsymbol{\theta}}\big|_{\boldsymbol{\theta}=\boldsymbol{\theta}^{t}}, \label{surrogate_gradient} 
\end{equation}
where $\{\varrho^t\}$ is a sequence to be properly chosen and the expression of $\frac{\partial g(\boldsymbol{\theta},\{\mathbf{P}_k^*,\mathbf{F}_{l}^*\};\mathcal{H}^t(m))}{\partial \boldsymbol{\theta}}$ is omitted here. 

Subsequently, we aim to solve the approximated problem at the $t$-th frame, which is given by \vspace{-0.2em}
\begin{equation} \vspace{-0.2em}
	\min_{\boldsymbol{\theta}} \quad \bar{f}^t(\boldsymbol{\theta}). 
\end{equation}
This is a convex quadratic problem and the optimal solution can be readily derived as \vspace{-0.2em}
\begin{equation} \vspace{-0.2em}
	\bar{\boldsymbol{\theta}}^t = \boldsymbol{\theta}^t-\frac{\mathbf{f}^t}{2\varpi}. \label{solution_to_surrogate} 
\end{equation}
Then, the long-term variable is updated as \vspace{-0.2em}
\begin{equation}  \vspace{-0.2em}
	\boldsymbol{\theta}^{t+1}=(1-\gamma^t)\boldsymbol{\theta}^{t} + \gamma^t\bar{\boldsymbol{\theta}}^t,  \label{theta_update} 
\end{equation} 
where $\{\gamma^t\}$ denotes a sequence of parameters and the convergence can be guaranteed if we choose $\varrho^t$ and $\gamma^t$ by following the conditions: $\lim_{t\rightarrow \infty} \varrho^t = 0, \sum_{t} \varrho^t  = \infty, \sum_{t} (\varrho^t)^2  < \infty,
\lim_{t\rightarrow \infty} \gamma^t = 0, \sum_{t} \gamma^t  = \infty, \sum_{t} (\gamma^t)^2  < \infty$ and $\lim_{t\rightarrow \infty} \frac{\gamma^t}{\varrho^t} = 0$~\cite{cssca,CRAN}. The proposed long-term passive beamforming design algorithm is summarized in \textbf{Algorithm 2}.  \vspace{-0.2em}

\begin{remark}
The proposed SSCA-based algorithm can be guaranteed to converge to a stationary solution of $\mathcal{P}4$~\cite{cssca}. Moreover, combining the convergence property of the BCD-type short-term active beamforming algorithm \cite{CRAN}, the proposed overall mixed-timescale joint active and passive beamforming algorithm converges to a stationary point of $\mathcal{P}1$. 
\end{remark} \vspace{-0.4em}

\begin{figure}[h]
	\centering
	\scalebox{0.60}{\includegraphics{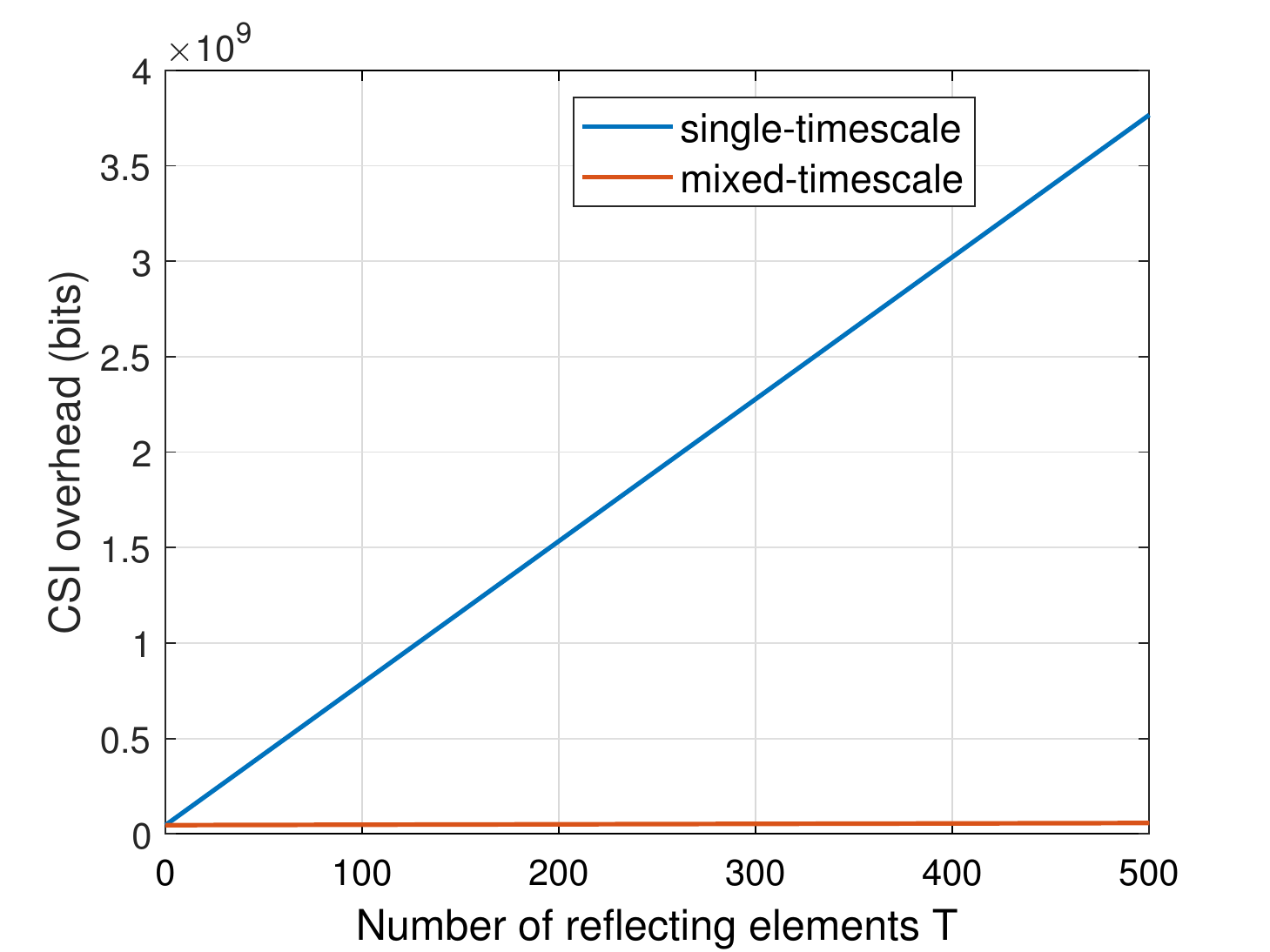}}
	\caption{CSI overhead versus the number of reflecting elements.}\label{fig:CSI_overhead} \vspace{-0.4em}
\end{figure}

For the single-timescale algorithm, the required CSI signaling bits in a coherence time block is given by $Q_s = qT_s(N_\text{U}KM_\text{U}+N_\text{D}LM_\text{D}+KLM_\text{U}M_\text{D}+T(N_\text{U}+N_\text{D}+2KM_\text{U}+2LM_\text{D}-3))$~\cite{IRS_CM2}, where $q$ is the quantization bits for each element of CSI matrices and $T_s$ denotes the number of time slots in a coherence time block while that of the mixed-timescale scheme is given by $Q_m = qT_s(N_\text{U}KM_\text{U}+N_\text{D}LM_\text{D}+KLM_\text{U}M_\text{D}) + qA_sT(N_\text{U}+N_\text{D}+2KM_\text{U}+2LM_\text{D}-3)$. Fig.~\ref{fig:CSI_overhead} compares the single-timescale scheme and the proposed mixed-timescale in terms of CSI overhead, where we set $q = 8, T_s = 10000$~\cite{FDrelay2}, $A_s = 30, K = L = 2, N_\text{U} = N_\text{D} = 32, M_\text{U} = M_\text{D} = 4$. From the figure, our proposed mixed-timescale algorithm can significantly reduce the CSI overhead, especially when $T$ is large.

\begin{algorithm}[t]\caption{Proposed SSCA-based algorithm for the long-term passive beamforming design.}
	\begin{algorithmic}[1]
		\footnotesize
		\begin{small}
			\STATE Initialize $\boldsymbol{\theta}^0$ with a feasible point. Select a proper sequence for $\{\varrho^t\}$ and $\{\gamma^t\}$. Set an appropriate value for $\varpi$ and let $t=0$.
			\REPEAT
			\STATE Randomly select $B$ samples $\{\mathcal{H}^t(m)\}_{m=\{1,...,B\}}$ from the collection of full CSI samples. Compute the surrogate function \eqref{surrogate function} based on \eqref{surrogate_gradient}.
			\STATE Obtain the optimal solution via \eqref{solution_to_surrogate}.
			\STATE Update $\boldsymbol{\theta}^t$ based on \eqref{theta_update}.
			\STATE Update the iteration number $t=t+1$.
			\UNTIL the convergence condition is satisfied or the maximum number of iterations is reached.
		\end{small}
	\end{algorithmic}
\end{algorithm}

\begin{figure*}[!t]
	\centering
	\scalebox{0.85}{\includegraphics{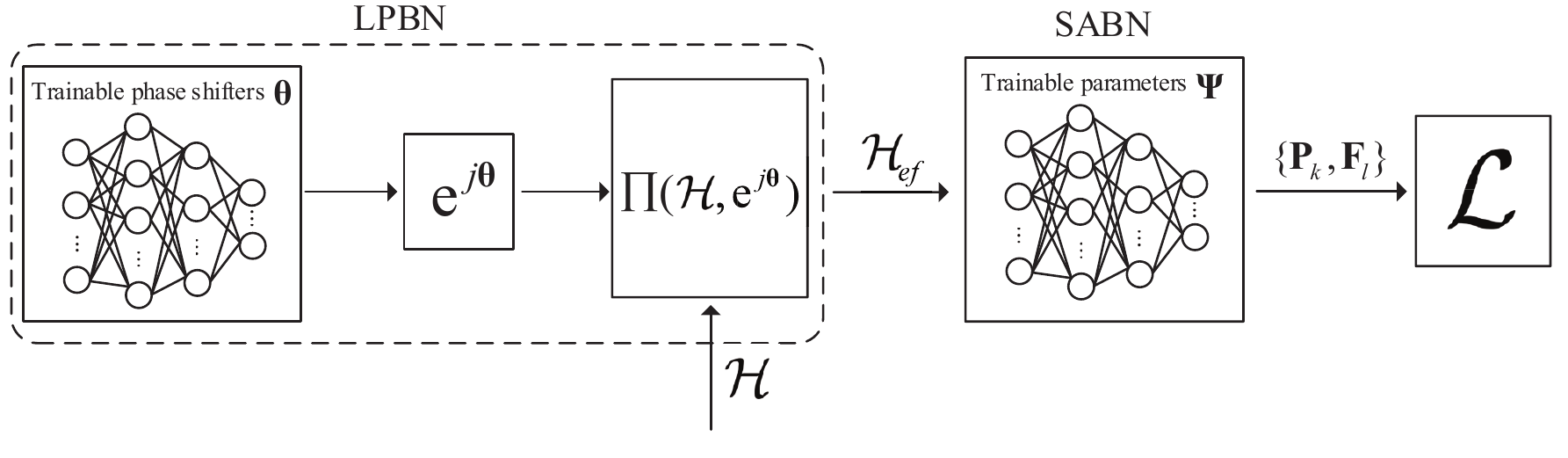}}
	\caption{Architecture of the proposed deep-unfolding NN consisting of the LPBN and SABN.}\label{fig:joint_structure}
\end{figure*}
\begin{figure*}[!t]
	\centering
	\scalebox{0.93}{\includegraphics{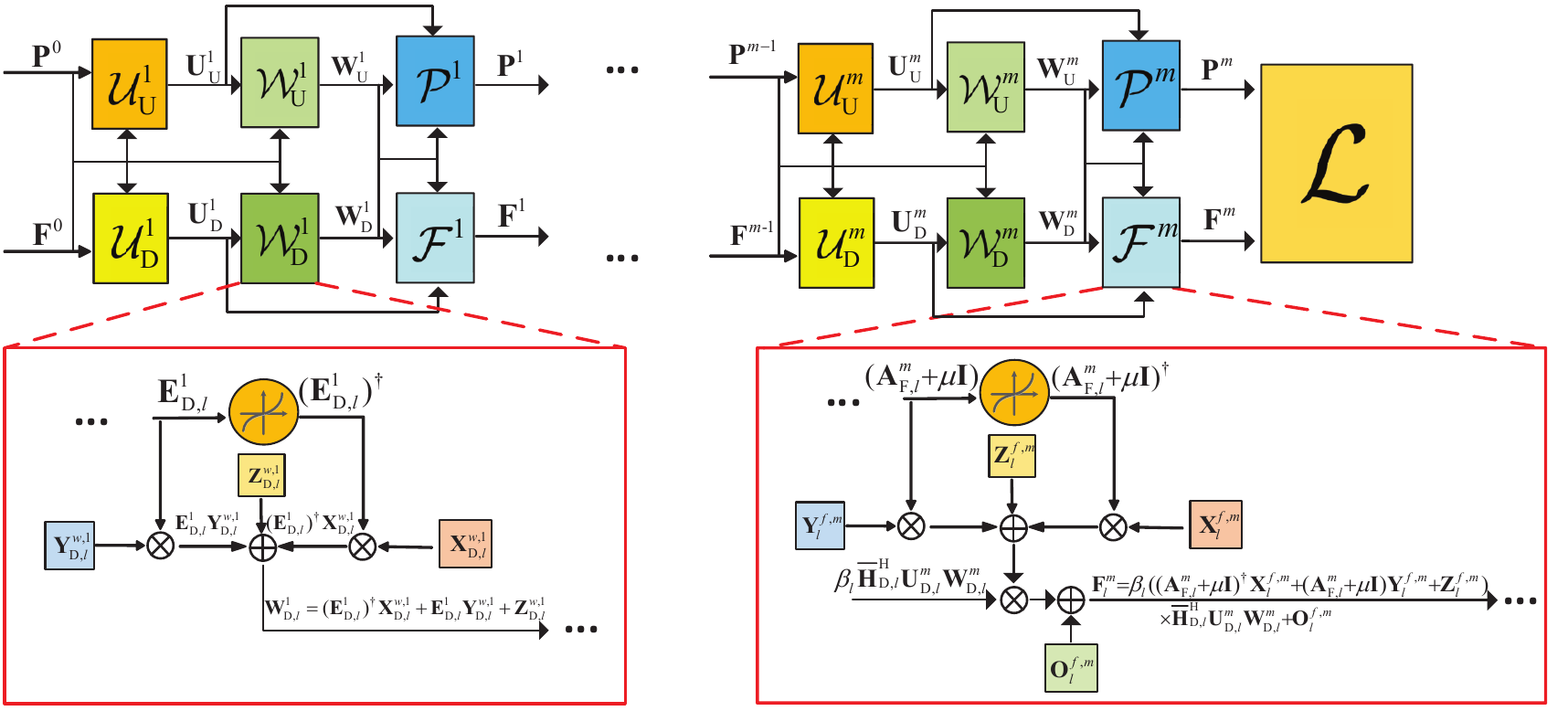}}
	\caption{Structure of the SABN.}\label{fig:short_unfolding} \vspace{-0.9em}
\end{figure*}

\section{Deep-Unfolding Beamforming}

In this section, we introduce the proposed deep-unfolding NN that unfolds the SSCA-based mixed-timescale beamforming algorithm. 
\vspace{-0.6em}
\subsection{Architecture of Deep-Unfolding NN}
The framework of our proposed deep-unfolding NN is shown in Fig. \ref{fig:joint_structure}. It consists of a LPBN and a SABN, which corresponds to the long-term passive beamforming algorithm and short-term active beamforming algorithm, respectively.

\subsubsection{Forward Propagation} 
In the forward propagation stage, the full CSI samples, $\mathcal{H}$, are first input into the LPBN, and then the effective CSI samples, $\mathcal{H}_{ef}$, are output. Note that we set the phase of the IRS passive beamforming vector $\boldsymbol{\theta}$ as the learnable parameter of LPBN and the operation $e^{j(\cdot)}$ ensures that the unit-modulus constraint is satisfied. Moreover, the function that computes the effective channels is given by \vspace{-0.1em}
\begin{equation} \vspace{-0.1em}
	\begin{split}
	\mathcal{H}_{ef} &= \Pi(\mathcal{H},e^{j\boldsymbol{\theta}})\triangleq\{\mathbf{\bar{H}}_{\text{U}, k}= \mathbf{H}_{\text{U},k}+\mathbf{V}_\text{U}\mathbf{\Phi}\mathbf{G}_{\text{U},k},\mathbf{\bar{H}}_{\text{D},l}= \\& \mathbf{H}_{\text{D},l}+\mathbf{G}_{\text{D},l}\mathbf{\Phi}\mathbf{V}_\text{D}, \mathbf{\bar{J}}_{k,l}= \mathbf{J}_{k,l}+\mathbf{G}_{\text{D,l}}\mathbf{\Phi}\mathbf{G}_{\text{U},k},\tilde{\mathbf{H}} = \tilde{\mathbf{H}}\},
	\end{split}
\end{equation}
where $\mathbf{\Phi} = \textrm{Diag}(e^{j\boldsymbol{\theta}})$. Then, the effective CSI samples pass through the SABN that outputs the active beamforming matrices $\mathbf{P}_k$ and $\mathbf{F}_l$. The detailed structure of the SABN will be introduced in Section \ref{ArchiSABN}. Denote $\mathcal{F}(\cdot)$ as the whole forward propagation stage of our proposed deep-unfolding NN, that is,
\begin{equation}
	\{\mathbf{P}_k,\mathbf{F}_l\} = \mathcal{F}(\{\boldsymbol{\theta},\mathbf{\Psi}\};\mathcal{H}),
\end{equation}    
where $\boldsymbol{\theta}$ and $\mathbf{\Psi}$ are the learnable parameters of the LPBN and SABN, respectively,  and $\{\mathbf{P}_k,\mathbf{F}_l\}$ are the output active beamforming matrices of the deep-unfolding NN.

\subsubsection{Loss Function} Since we aim to maximize the weighted sum-rate of the system, the loss function of the deep-unfolding NN can be expressed as
\begin{equation}
	\mathcal{L}(\{\boldsymbol{\theta},\mathbf{\Psi}\};\mathcal{H}) \triangleq g(\boldsymbol{\theta},\mathcal{F}(\{\boldsymbol{\theta},\mathbf{\Psi}\};\mathcal{H});\mathcal{H}), \label{loss_function}
\end{equation}
where $g(\cdot)$ is the sum-rate function defined in \eqref{gfunction}. 

\subsubsection{Back Propagation} In the back propagation stage, the gradients of the learnable parameters, $\frac{\partial \mathcal{L}(\{\boldsymbol{\theta},\mathbf{\Psi}\};\mathcal{H})}{\partial \boldsymbol{\theta}}$ and $\frac{\partial \mathcal{L}(\{\boldsymbol{\theta},\mathbf{\Psi}\};\mathcal{H})}{\partial \boldsymbol{\Psi}}$, are computed based on the chain rule. 

\subsubsection{Update of Learnable Parameters}  We update $\{\boldsymbol{\theta},\mathbf{\Psi}\}$ based on the gradients of the learnable parameters. Specifically, in the $t$-th round of the learning process, the learnable parameters are updated as 
\begin{equation}
	\boldsymbol{\theta}^{t+1} = \boldsymbol{\theta}^{t} - \eta  \frac{\partial \mathcal{L}(\{\boldsymbol{\theta},\mathbf{\Psi}\};\mathcal{H}^t)}{\partial \boldsymbol{\theta}},
\end{equation}
\begin{equation}
	\mathbf{\Psi}^{t+1} = \mathbf{\Psi}^{t} - \eta \frac{\partial \mathcal{L}(\{\boldsymbol{\theta},\mathbf{\Psi}\};\mathcal{H}^t)}{\partial \boldsymbol{\Psi}},
\end{equation}  
where $\eta$ denotes the learning rate. Since the LPBN is an approximation of the SSCA-based algorithm, we can also update $\boldsymbol{\theta}$ based on~\eqref{surrogate_gradient},~\eqref{solution_to_surrogate}, and~\eqref{theta_update}, correspondingly.
 
\subsection{Structure of the SABN} \label{ArchiSABN}
The LPBN sets the IRS passive beamforming vector $\boldsymbol{\theta}$ as a learnable parameter and its forward propagation is to compute effective channels $\mathcal{H}_{ef}$. 
In this subsection, we introduce the detailed structure of the SABN, which unfolds Algorithm \ref{BCDtype} into a layer-wise structure. 

We first define a novel element-wise non-linear operation that takes the reciprocal of each element in the diagonal of matrix $\mathbf{A}$ while setting the non-diagonal elements to be $0$, i.e., denoted as $\mathbf{A}^{\dagger}$. We take a $3\times 3$ matrix as an example, 

\begin{equation}       
	\mathbf{A} = \left[                
	\begin{array}{ccc}  
		a_{11} & a_{12} & a_{13}\\  
		a_{21} & a_{22} & a_{23}\\
		a_{31} & a_{32} & a_{33}\\
	\end{array}
	\right],   
	\quad
	\mathbf{A}^{\dagger} = \left[                 
	\begin{array}{ccc}  
		\frac{1}{a_{11}} & 0 & 0\\  
		0 & \frac{1}{a_{22}}  & 0\\
		0 & 0 & \frac{1}{a_{33}} \\
	\end{array}
	\right].                
\end{equation}
Note that $\mathbf{A}^{-1}=\mathbf{A}^{\dagger}$ when $\mathbf{A}$ is a diagonal matrix. We observe that the diagonal elements of the matrices are much larger than the off-diagonal elements in the proposed BCD-type algorithm. Hence, $\mathbf{A}^{\dagger}$ is a good estimation of $\mathbf{A}^{-1}$. 
Since solving the matrix inversion $\mathbf{A}^{-1}$ requires high computational complexity, we approximate it by employing the combination of the following two architectures with lower complexity: 
(i) $\mathbf{A}^{\dagger}\mathbf{X}$ with the element-wise non-linear function $\mathbf{A}^{\dagger}$ and learnable parameter $\mathbf{X}$; 
(ii) By recalling the first-order Taylor expansion of $\mathbf{A}^{-1}$ at $\mathbf{A}_0$: $\mathbf{A}^{-1}=2\mathbf{A}_{0}^{-1}-\mathbf{A}_{0}^{-1}\mathbf{A}\mathbf{A}_{0}^{-1}$, we apply $\mathbf{A}\mathbf{Y}
+ \mathbf{Z}$ with learnable parameters $\mathbf{Y}$ and $\mathbf{Z}$.
Note that the learnable parameters $\mathbf{X}$, $\mathbf{Y}$, and $\mathbf{Z}$ are introduced to improve the performance of the deep-unfolding NN. 

Thus, we apply  $\mathbf{A}^{\dagger}\mathbf{X}+\mathbf{A}\mathbf{Y}+\mathbf{Z}$ to approximate matrix inversion $\mathbf{A}^{-1}$.
Note that $\mathbf{\Xi}^m \triangleq \{ \mathbf{X}_{{\rm U},k}^{u,m}, \mathbf{Y}_{{\rm U},k}^{u,m},  \mathbf{Z}_{{\rm U},k}^{u,m} \}$$\cup$$\{ \mathbf{X}_{{\rm D},l}^{u,m}$\\
$, \mathbf{Y}_{{\rm D},l}^{u,m}, \mathbf{Z}_{{\rm D},l}^{u,m} \}$$\cup$$\{ \mathbf{X}_{{\rm U},k}^{w,m}, \mathbf{Y}_{{\rm U},k}^{w,m},\mathbf{Z}_{{\rm U},k}^{w,m} \}$$\cup$$\{ \mathbf{X}_{{\rm D},l}^{w,m}, \mathbf{Y}_{{\rm D},l}^{w,m}, \mathbf{Z}_{{\rm D},l}^{w,m} \}$\\
$\cup$$\{ \mathbf{X}_{k}^{p,m}, \mathbf{Y}_{k}^{p,m}, \mathbf{Z}_{k}^{p,m} \}$$\cup$$\{ \mathbf{X}_{l}^{f,m}, \mathbf{Y}_{l}^{f,m}, \mathbf{G}_{\text{D},l}^{f,m} \}$\footnote{Note that here the subscripts $\text{U}$ and $\text{D}$ correspond to the uplink and downlink, respectively. Subscripts $k$ and $l$ represent the $k$-th UL user and the $l$-th DL user, respectively. Superscripts $u$, $w$, $p$ and $f$ denote the corresponding unfolding matrices and superscript $m$ denotes the layer index.} are introduced learnable parameter sets to approximate the inversion of matrix variables $\mathbf{U}_{{\rm U},k}^{m}$, $\mathbf{U}_{{\rm D},l}^{m}$, $\mathbf{W}_{{\rm U},k}^{m}$, $\mathbf{W}_{{\rm D},l}^{m}$, $\mathbf{P}_{k}^{m}$, and $\mathbf{F}_{l}^{m}$ in the $m$-th layer, respectively, and $\{ \mathbf{O}_{{\rm U},k}^{u,m}, \mathbf{O}_{{\rm D},l}^{u,m}, \mathbf{O}_{k}^{p,m}, \mathbf{O}_{l}^{f,m} \}$ denote the learnable offsets. The architecture of the SABN is designed as:
\begin{figure*}[t]
\begin{subequations}  \label{network}
	\begin{eqnarray}
		& & \!\!\!\!\! \mathbf{U}_{{\rm U},k}^{m} = \bigg( (\mathbf{A}_{{\rm U},k}^{m-1})^{\dagger}\mathbf{X}_{{\rm U},k}^{u,m} + \mathbf{A}_{{\rm U},k}^{m-1}\mathbf{Y}_{{\rm U},k}^{u,m}
		+ \mathbf{Z}_{{\rm U},k}^{u,m}  \bigg) \bar{\mathbf{H}}_{{\rm U},k}\mathbf{P}_{k}^{m-1} + \mathbf{O}_{{\rm U},k}^{u,m}, \label{UU} \\
		& & \!\!\!\!\! \mathbf{U}_{{\rm D},l}^{m} = \bigg( (\mathbf{A}_{{\rm D},l}^{m-1})^{\dagger}\mathbf{X}_{{\rm D},l}^{u,m} + \mathbf{A}_{{\rm D-1},l}^{m}\mathbf{Y}_{{\rm D},l}^{u,m}
		+ \mathbf{Z}_{{\rm D},l}^{u,m}  \bigg)
		\bar{\mathbf{H}}_{{\rm D},l}\mathbf{F}_{l}^{m-1} + \mathbf{O}_{{\rm D},l}^{u,m}, \label{UD} \\
		& & \!\!\!\!\! \mathbf{W}_{{\rm U},k}^{m} = (\mathbf{E}_{{\rm U},k}^{m})^{\dagger}\mathbf{X}_{{\rm U},k}^{w,m} + \mathbf{E}_{{\rm U},k}^{m}\mathbf{Y}_{{\rm U},k}^{w,m}
		+ \mathbf{Z}_{{\rm U},k}^{w,m},  \label{WU} \\
		& & \!\!\!\!\! \mathbf{W}_{{\rm D},l}^{m} = (\mathbf{E}_{{\rm D},l}^{m})^{\dagger}\mathbf{X}_{{\rm D},l}^{w,m} + \mathbf{E}_{{\rm D},l}^{m}\mathbf{Y}_{{\rm D},l}^{w,m}
		+ \mathbf{Z}_{{\rm D},l}^{w,m},  \label{WD}  \\
		& & \!\!\!\!\! \mathbf{P}_{k}^{m} \!\!=\!\! \alpha_k
		\bigg( (\mathbf{A}_{\text{P},k}^{m} \!+\! \lambda_k^m\mathbf{I})^{\dagger}\mathbf{X}_{k}^{p,m} \!+\! (\mathbf{A}_{\text{P},k}^{m} \!+\! \lambda_k^m\mathbf{I})\mathbf{Y}_{k}^{p,m} \!+\! \mathbf{Z}_{k}^{p,m}  \bigg)
		\bar{\mathbf{H}}_{{\rm U}, k}^{\rm H}\mathbf{U}_{{\rm U},k}^{m}\mathbf{W}_{{\rm U},k}^{m} \!+\! \mathbf{O}_{k}^{p,m}, \label{PK} \\
		& & \!\!\!\!\! \mathbf{F}_{l}^{m} \!\!=\!\! \beta_l  \bigg( (\mathbf{A}_{\text{F},l}^{m} \!+\! \mu^m \mathbf{I})^{\dagger}\mathbf{X}_{l}^{f,m} \!+\! (\mathbf{A}_{\text{F},l}^{m} \!+\! \mu^m \mathbf{I})\mathbf{Y}_{l}^{f,m} \!+\! \mathbf{G}_{\text{D},l}^{f,m}  \bigg) \bar{\mathbf{H}}_{{\rm D}, l}^{\rm H}\mathbf{U}_{{\rm D},l}^{m}\mathbf{W}_{{\rm D},l}^{m} \!+\! \mathbf{O}_{l}^{f,m}. \label{FL}
	\end{eqnarray}
\end{subequations} \vspace{-1em}
\end{figure*}

The architecture of the SABN is presented in Fig.~\ref{fig:short_unfolding}, where $\mathcal{U}^{m}_{\rm U}$, $\mathcal{U}^{m}_{\rm D}$, $\mathcal{W}^{m}_{\rm U}$, $\mathcal{W}^{m}_{\rm D}$, $\mathcal{P}^{m}$, and $\mathcal{F}^{m}$ represent the layers of the deep-unfolding NN, i.e., \eqref{UU}-\eqref{FL}. In addition, the Lagrange multipliers, $\lambda_k^m$ and $\mu^m$, are also set as learnable parameters. Hence, the learnable parameters of the SABN can be denoted as $\mathbf{\Psi} \triangleq \bigcup_{m=1}^{I_u}\mathbf{\Xi}^m\cup\{\mathbf{O}_{{\rm U},k}^{u,m}, \mathbf{O}_{{\rm D},l}^{u,m}, \mathbf{O}_{k}^{p,m}, \mathbf{O}_{l}^{f,m},\lambda_k^m,\mu^m\}$, where $I_u$ is the number of layers.  Moreover, to avoid gradient explosion and ensure that the power constraints \eqref{shorttransmitpower} and \eqref{shorttransmitpower2} are satisfied, we scale each $\mathbf{F}_{l}^{m}$ as $\frac{ \sqrt{P_{AP}} }{ \big( \sum\limits_{l} \textrm{Tr}(\mathbf{F}_{l}^{m}(\mathbf{F}_{l}^{m})^{H}) \big)^{\frac{1}{2}} }\mathbf{F}_{l}^{m}$. Note that $\mathbf{P}_{k}^{m}$ can be scaled in the same way. The output layer is a single-layer BCD iteration as~\cite{unfolding8}.\vspace{-0.3em}
\begin{remark}
	Let us investigate the connection between the deep-unfolding NN and the SSCA-based algorithm. First, it is obvious that the SABN has a similar structure with the BCD-type short-term active beamforming algorithm. However, the SABN introduces learnable parameters to approximate the matrix inversion. Second, regarding the relation between the LPBN and the SSCA-based long-term passive beamforming algorithm, let us focus on the surrogate functions of these two approaches. Specifically, the surrogate function of the SSCA-based algorithm is constructed based on the sample gradient $\frac{\partial g(\boldsymbol{\theta},\{\mathbf{P}_k^*,\mathbf{F}_{l}^*\};\mathcal{H})}{\partial \boldsymbol{\theta}}\big|_{\boldsymbol{\theta}=\boldsymbol{\theta}^{t}}$, while that of the deep-unfolding NN is constructed based on
	\begin{equation}
		\begin{split}			
		\frac{\partial \mathcal{L}(\{\boldsymbol{\theta},\mathbf{\Psi}\};\mathcal{H})}{\partial \boldsymbol{\theta}}\big|_{\boldsymbol{\theta}=\boldsymbol{\theta}^{t}} &= \frac{\partial g(\boldsymbol{\theta},\mathcal{F}(\{\boldsymbol{\theta},\mathbf{\Psi}\};\mathcal{H});\mathcal{H})}{\partial \boldsymbol{\theta}}\big|_{\boldsymbol{\theta}=\boldsymbol{\theta}^{t}} \\
		& = \frac{\partial g(\boldsymbol{\theta},\mathcal{F}(\{\boldsymbol{\theta}^t,\mathbf{\Psi}\};\mathcal{H});\mathcal{H})}{\partial \boldsymbol{\theta}}\big|_{\boldsymbol{\theta}=\boldsymbol{\theta}^{t}}\\
		&+(\frac{\partial g}{\partial \mathcal{F}})^T\frac{\partial \mathcal{F}(\{\boldsymbol{\theta},\mathbf{\Psi};\mathcal{H}\}}{\partial \boldsymbol{\theta}}\big|_{\boldsymbol{\theta}=\boldsymbol{\theta}^{t}}.	
	    \end{split} \label{gradient_comparsion}
	\end{equation} 
    The first term in the last row of \eqref{gradient_comparsion} is  the same as the sample gradient of the SSCA-based algorithm except that the active beamforming matrices are obtained by the network $\mathcal{F}(\{\boldsymbol{\theta}^t,\mathbf{\Psi}\};\mathcal{H})$ instead of the BCD-type algorithm $\{\mathbf{P}_k^*,\mathbf{F}_{l}^*\}$. The second term that represents the gradient of the network only appears in the deep-unfolding NN and is not included in the SSCA-based algorithm. This is because the jointly designed structure of the proposed deep-unfolding NN ties the long-term IRS passive beamforming matrix and the short-term active beamforming matrices more tightly. \vspace{-0.2em}
\end{remark}

\vspace{-0.8em}
\subsection{Black-box NN for Beamforming Design}
\begin{figure*}[!t]
	\centering
	\scalebox{0.5}{\includegraphics{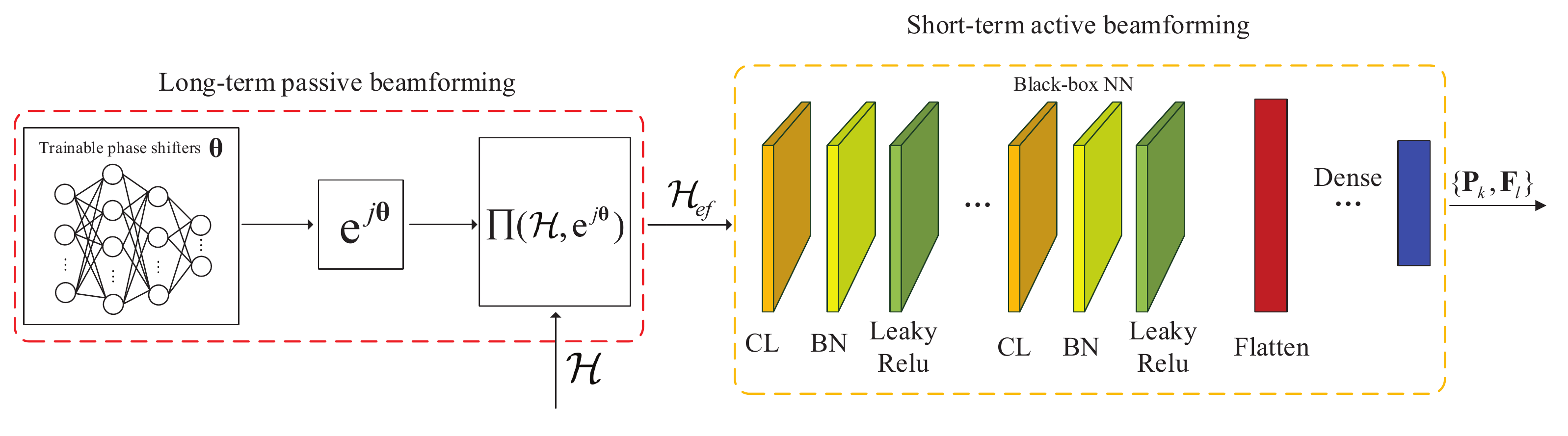}}
	\caption{Structure of the black-box NN.}\label{fig:blackbox} \vspace{-0.8em}
\end{figure*}

	In this subsection, we propose a black-box NN for comparison. As shown in Fig. \ref{fig:blackbox}, the black-box NN also consists of two parts. The long-term passive beamforming part is the same as that of our proposed deep-unfolding NN. The short-term active beamforming part is comprised of conventional black-box layers, such as convolutional layers (CLs) and fully connected layers (FCLs). Specifically, full channel samples $\mathcal{H}$ first pass through the long-term passive beamforming part and it outputs the effective channels $\mathcal{H}_{ef}$. Then, the effective channels enter the CL, the batch normalization (BN) layer, and the non-linear function in serial and this process repeats for a number of times. Subsequently, the outputs are flattened and pass through several FCLs followed by a non-linear function. In particular, we adopt Leaky ReLU as the non-linear function, i.e., $y=\max\{x,x/a\}$, where $a>1$ is a constant. The final outputs of the black-box NN are the active beamforming matrices $\mathbf{P}_k$ and $\mathbf{F}_l$, and we scale them to satisfy the power constraint. The weighted sum-rate in \eqref{loss_function} is employed as the loss function. 

\vspace{-0.8em}
\subsection{Computational Complexity} \vspace{-0.1em}
In this subsection, we analyze the computational complexity of the proposed SSCA-based optimization algorithm, the proposed deep-unfolding NN, the benchmark black-box NN, and the conventional single-timescale algorithm.

The computational complexity of the SSCA-based algorithm is dominated by the short-term BCD-type active beamforming algorithm, which is given by 
$\mathcal{O}\big( I_m \big( K(N_r^3 + M_U^3) + L(N_t^3 +M_D^3) + K^2N_r^2M_U + L^2N_t^2M_D \big) \big)$, where $I_m$ denotes the number of iterations.

The computational complexity of the deep-unfolding NN is given by $\mathcal{O}\big( I_u \big( K(N_r^{2.37} + M_U^{2.37}) + L(N_t^{2.37} +M_D^{2.37}) + K^2N_r^2M_U + L^2N_t^2M_D \big) \big)$, where $I_u \ll I_m$ is the number of layers. Compared to the complexity of the iterative BCD-type algorithm, the deep-unfolding NN has much lower complexity for the following two reasons: (i) The number of layers in the deep-unfolding NN is much smaller than that of the BCD-type algorithm; (ii) The iterative BCD-type algorithm involves the matrix inversion with complexity $\mathcal{O}(N^3)$
while the deep-unfolding NN simply requires matrix multiplication with complexity $\mathcal{O}(N^{2.37})$.

The computational complexity of the black-box NN is $\mathcal{O}\big( \sum_{l=1}^{L_{c}-1}Q_l^2S_l^2C_{l-1}C_l + C_{L_c}Q_{L_{c}}F_1 + \sum_{l=2}^{L_{f}}F_{l-1}F_{l} + F_{L_f}(KM_\text{U}D_\text{U} + LN_tD_\text{D}) \big)$, where $L_c$ is the number of CLs, $L_f$ is the number of FCLs, $S_l$ represents the size of the convolutional kernel, $C_l$ denotes the number of channels in the $l$-th CL, $Q_l$ denotes the output size of the $l$-th CL, which depends on the input size, padding number, and stride, and $F_l$ is the output size of the $l$-th FCL.

Moreover, we analyze the computational complexity of the single-timescale algorithm for comparison. Specifically, the single-timescale algorithm collects real-time high-dimensional full CSI samples and optimizes the active beamforming matrices and the IRS passive beamforming matrix employing a BCD-type algorithm in each time slot. The procedure for updating the active beamforming matrices is given in Algorithm \ref{BCDtype}. Regarding the optimization of the IRS passive beamforming matrix, a one-iteration BCD algorithm is adopted~\cite{oneiterationBCD}, whose computational complexity is given by $\mathcal{O}(T^3)$. Hence, the overall computational complexity of the single-timescale algorithm is given by $\mathcal{O}\big( I_s \big( T^3 + K(N_r^3 + M_U^3) + L(N_t^3 +M_D^3) + K^2N_r^2M_U + L^2N_t^2M_D \big) \big)$. It is readily seen that our proposed mixed-timescale scheme can significantly reduce the computational complexity compared with the single-timescale algorithm since $T$ is generally large. 

\vspace{-0.3em}
\section{Simulation Results}

In this section, we present simulation results to evaluate the performance of our proposed algorithms. 
The simulation setting is shown in Fig.~\ref{fig:setting}. The AP is located at $(0 \,\text{m}, 0\,\text{m}, 0 \,\text{m})$ and the position of the IRS is $(0\, \text{m}, d_1 = 80\, \text{m}, 3\, \text{m})$. We consider $K = 2$ UL users and $L = 2$ DL users and they lie on the  corners of a square centered at $(0\,\text{m}, 80\, \text{m}, 0\, \text{m})$ with a side length $20\, \text{m}$. Both AP and users are equipped with uniform linear arrays (ULA) and the IRS is equipped with a uniform planar array (UPA). The distance-dependent path loss is modeled as $L(d) = C_0(\frac{d_{link}}{D_{0}})^{-a}$, where $C_0$ is the path loss at the reference distance $D_0 =1\, \text{m}$, $d_{link}$ represents the individual link distance, and $a$ denotes the path loss exponent.  As for the small-scale fading, we assume the Rician fading channel model, which is given by 
\begin{equation}
	\mathbf{H} = \sqrt{\frac{b_{}}{1+b_{}}}\mathbf{H}^{\text{Los}} + \sqrt{\frac{1}{1+b_{}}}\mathbf{H}_{}^{\text{NLos}},
\end{equation} 
where $b_{}$ is the Rician factor, and $\mathbf{H}^{\text{Los}}$ and $\mathbf{H}^{\text{NLos}}$ represent the deterministic line-of-sight (LoS) and random Rayleigh fading components, respectively. In particular, we let $a_{AI}$, $a_{Au}$, $a_{{Iu}}$, and $a_{uu}$ denote the path loss exponents of the AP-IRS link, AP-user link, IRS-user link, and user-user link, respectively, and let $b_{AI}$, $b_{Au}$, $b_{Iu}$, and $b_{uu}$ represent the Rician factor of these links, respectively. The residual SI channel
matrix $\tilde{\mathbf{H}}$ is generated based on the model described in \cite{SIchannel}, and the average power of the SI channel is denote by $\sigma_{SI}^2$. The system parameters are set as follows unless otherwise stated: $N_t = N_r = N =32$, $M_{\text{U},k} = M_{\text{D},l} = 4, D_{\text{U},k} = D_{\text{D},l} = 4, \forall k,l$, $T = 200$, $\sigma_{\text{U}}^2 = \sigma_{\text{D},l}^2 = -76 $ dBm$, \forall l$, $\sigma_{SI}^2 = -60 $ dB,  $P_{\text{U},k} = 24$ dBm$, \forall k$, $P_{AP} = 44$ dBm, $\alpha_{k} = \beta_{l} = 1, \forall k,l$, $a_{AI} = 2.4$, $a_{Au} = 3.8$, $a_{Iu} = 2.2$, $a_{uu} = 3.0$, $b_{AI} = b_{Iu} = 3$ dB, $b_{Au} = -3$ dB, and $b_{uu} = 0$ dB. For the algorithm parameters, we set $I_{max} = 100$, $\delta = 10^{-4}$, $\varrho^t = \frac{10}{(10+t)^{0.6}}$, $\gamma^t = \frac{15}{15+t}$, $\varpi = 0.5$. The default layer number of the proposed deep-unfolding NN is $8$ and the learning rate is chosen as $0.001$. The black-box NN consists of $3$ CLs and $5$ FCLs, and we adopt the Adam optimizer with the same learning rate $0.001$. Moreover, the batch size of the SSCA-based algorithm, the deep-unfolding NN, and the black-box NN are all set as $5$. All the experiments are conducted on a desktop Intel CPU (i5-8400 with 6 cores) with 8GB RAM. The benchmarks are provided as follows:
\begin{figure}[!t]
	\centering
	\scalebox{0.66}{\includegraphics{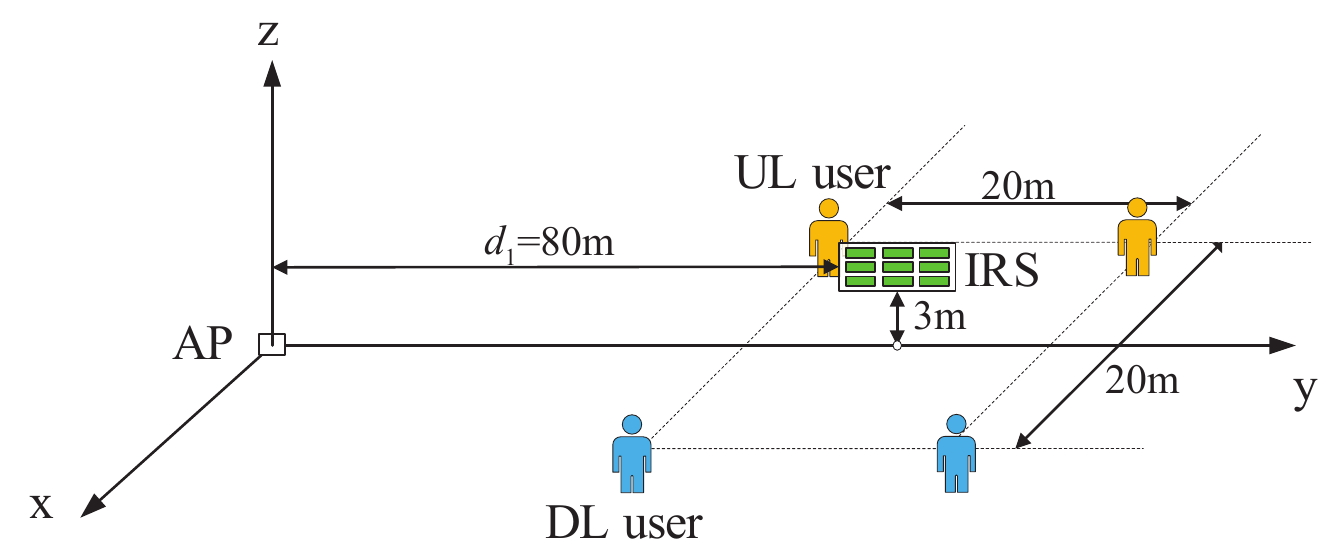}}
	\caption{Simulation setup.}\label{fig:setting}
	\vspace{-1em}
\end{figure}
\begin{itemize}
	\item SSCA: The proposed SSCA-based mixed-timescale joint active and passive beamforming algorithm.
	\item Deep-unfolding NN: The proposed deep-unfolding NN  introduced in Section IV.
	\item Black-box NN: The benchmark black-box NN.
	\item Full CSI: The single-timescale algorithm that collects high-dimensional full CSI in each time slot and optimizes the active beamforming matrices using Algorithm 1 and optimizes the IRS passive beamforming matrix using the one-iteration BCD algorithm.
	\item No IRS: The conventional scheme without IRS which directly employs Algorithm \ref{BCDtype} to optimize the active beamforming matrices.
	\item Random IRS: In this algorithm, the IRS passive beamforming matrix is randomly generated and Algorithm \ref{BCDtype} is adopted to optimize the active beamforming matrices. 
	\item HD: 
	The conventional HD scheme where the WMMSE algorithm is adopted for optimizing the active beamforming matrices and the IRS passive beamforming matrix is randomly generated.   
\end{itemize} 

\begin{figure}[t]
	\centering
	\subfloat[]{\label{fig:short-term_convergence}{\includegraphics[width=0.25\textwidth]{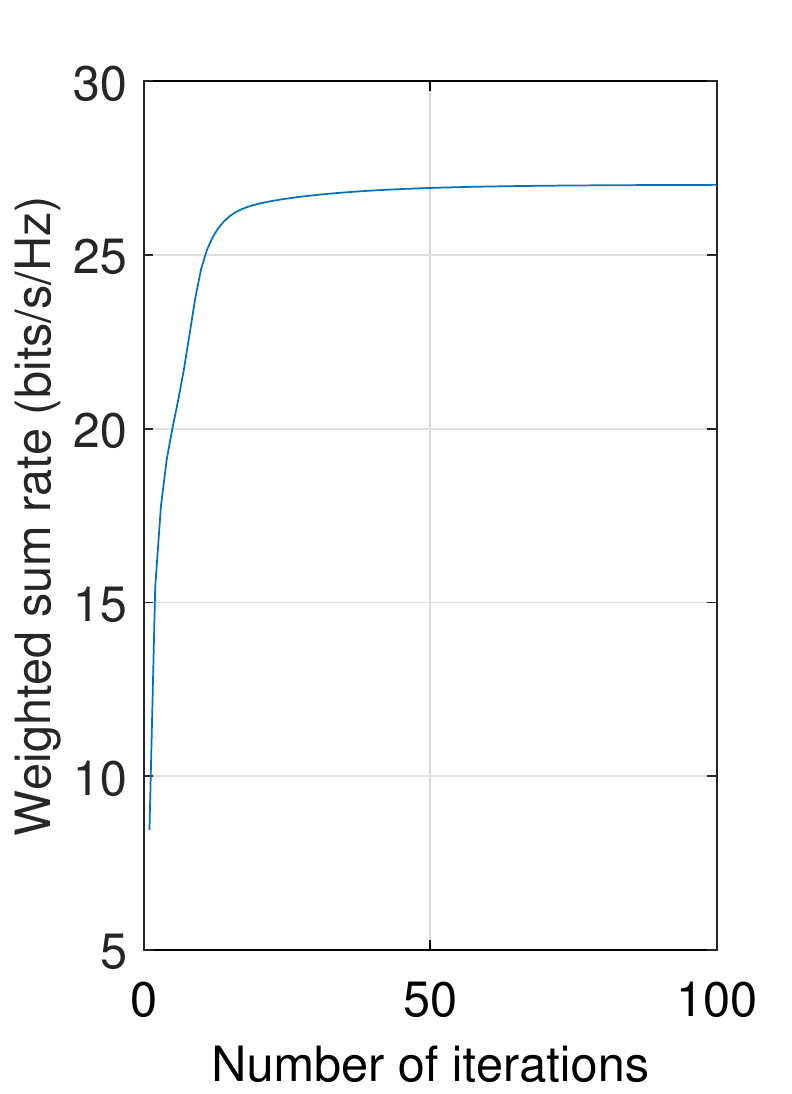}}}
	\subfloat[]{\label{fig:long-term_convergence}{\includegraphics[width=0.25\textwidth]{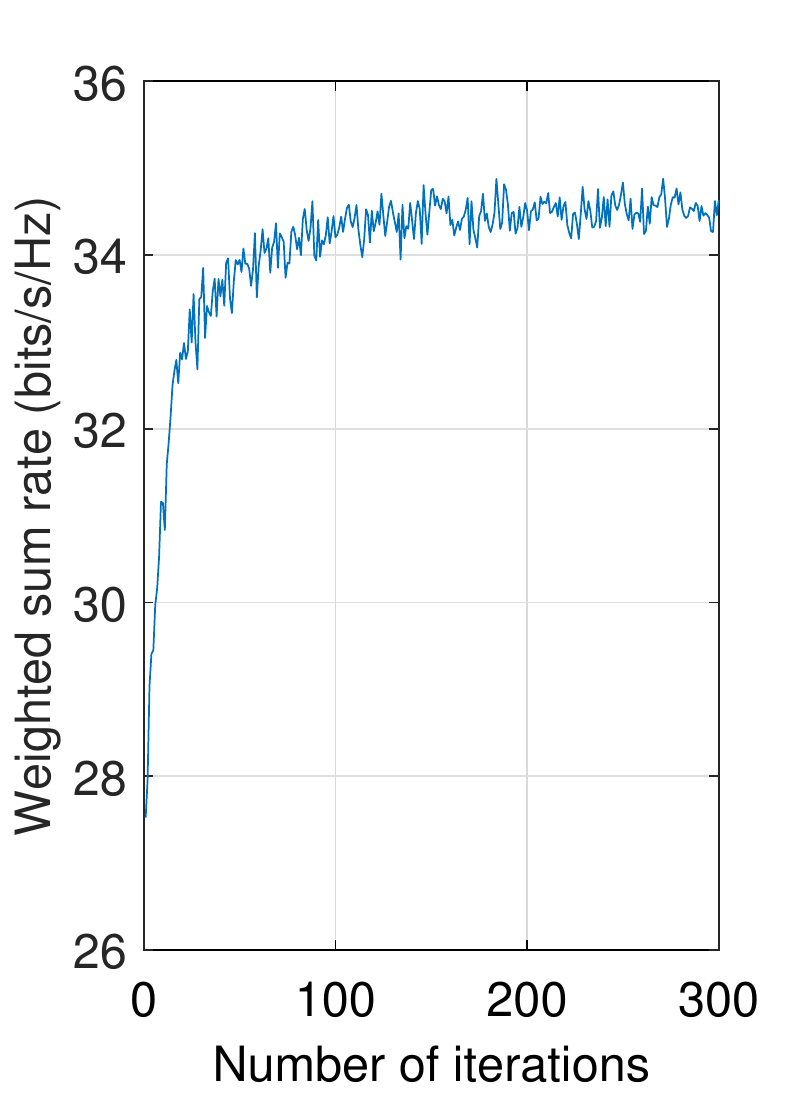}}}
	\caption{(a) Convergence performance of the BCD-type short-term active beamforming algorithm. (b) Convergence performance of the SSCA-based long-term passive beamforming algorithm.}
	\label{fig:fig3}
	\vspace{-0.4em}
\end{figure}

\begin{figure}[!t]
\centering
\scalebox{0.63}{\includegraphics{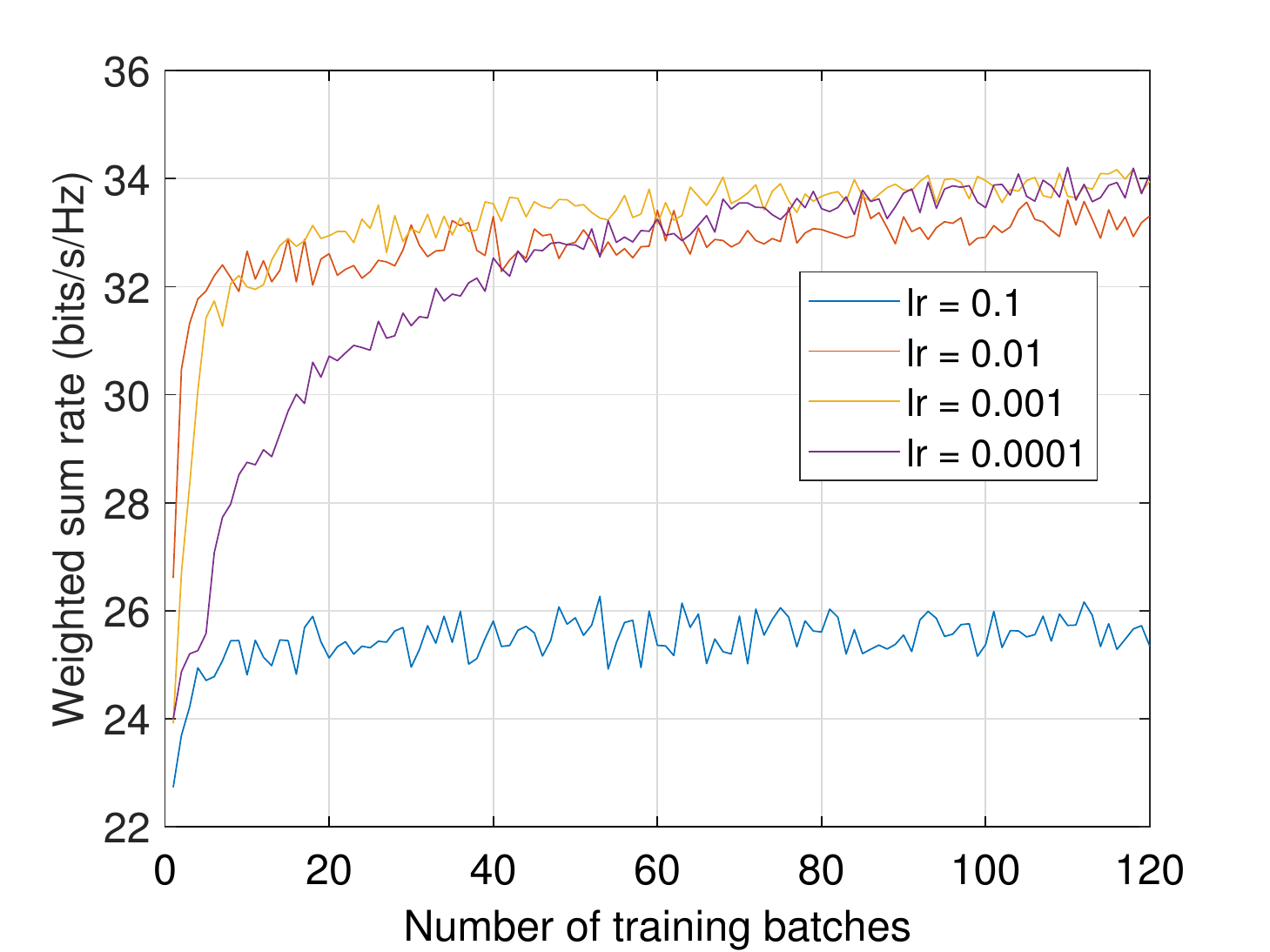}}
\caption{Convergence performance of the proposed deep-unfolding NN.}\label{fig:unfolding_convergence} \vspace{-0.8em}
\end{figure}	
	Fig.~\ref{fig:fig3}(\subref*{fig:short-term_convergence}) shows the convergence behavior of the proposed short-term active beamforming algorithm. It is observed that the BCD-type algorithm converges monotonically within 100 iterations. Fig.~\ref{fig:fig3}(\subref*{fig:long-term_convergence}) shows the value of the objective function versus the number of iterations for the SSCA-based long-term passive beamforming algorithm. From this figure, the weighted sum-rate converges within around 100 iterations. Moreover, we also observe some fluctuations of the objective function, which is due to the randomness of the sampled channels. Fig.~\ref{fig:unfolding_convergence} presents the impact of the learning rate on the convergence performance of the proposed deep-unfolding NN. As we can see, the sum-rate performance becomes better when the learning rate decreases from $0.1$ to $0.001$. When further decreasing the learning rate to $0.0001$, there is no evident performance improvement but the convergence speed significantly slows down. Hence, we choose $0.001$ as the learning rate in our test.
  
\begin{table*}[htbp]
	\centering
	\caption{Weighted sum-rate performance versus the number of collected/training samples.}
	\begin{tabular}{ccccccccc}
		\hline  
		Collected samples&5&10&15&20&25&30&35&40 \\ 
		\hline  
		SSCA&$88.79\%$&$93.73\%$&$97.02\%$&$98.32\%$&$99.57\%$&$99.92\%$&$100\%$&$100\%$ \\
		\hline 
	\end{tabular}
	 \vspace{0.4mm}
	 
	\begin{tabular}{ccccccccc}
		\hline
		Training samples &100&200&300&400&500&600&700&800\\ 
		\hline
		Deep-unfolding NN&$90.02\%$&$94.13\%$&$96.04\%$&$97.53\%$&$97.68\%$&$97.70\%$&$97.71\%$&$97.71\%$ \\
		\hline 
	\end{tabular}
		\vspace{0.4mm}
		
	\begin{tabular}{ccccccccc}
		\hline
		Training samples&500&1000&1500&2000&2500&3000&3500&4000 \\ 
		\hline
		Black-box NN&$75.03\%$&$79.59\%$&$83.20\%$&$85.52\%$&$86.27\%$&$86.42\%$&$86.62\%$&$86.64\%$ \\
       \hline
		\label{table:sample}
	\end{tabular} \vspace{-0.4em}
\end{table*}
\begin{table*}[htbp]
\centering
\caption{Weighted sum-rate performance versus the number of layers.}
\begin{tabular}{c|ccccccccc}
	\hline  
	Layers&2&3&4&5&6&7&8&9&10 \\ 
	\hline  
	SSCA&$78.02\%$&$80.01\%$&$83.08\%$&$85.41\%$&$86.55\%$&$87.34\%$&$88.46\%$&$89.35\%$&$90.21\%$ \\
	\hline
	Deep-unfolding NN&$90.76\%$&$92.56\%$&$95.70\%$&$96.31\%$&$97.57\%$&$97.69\%$&$97.72\%$&$97.71\%$&$97.72\%$ \\
	\hline
	Black-box NN&$86.09\%$&$86.23\%$&$86.50\%$&$86.83\%$&$86.18\%$&$86.81\%$&$86.21\%$&$86.64\%$&$86.17\%$ \\
	\hline 
\end{tabular} \vspace{0.1em}
\label{table:layer}
\end{table*}

Table~\ref{table:sample} shows the weighted sum-rate performance versus the number of collected/training samples. Note that the results are all normalized by a reference value, which is the weighted sum-rate of the SSCA-based algorithm that collects $100$ samples for optimizing $\boldsymbol{\theta}$. As we can see, the SSCA-based algorithm is very efficient and only about 30 samples are sufficient for it to learn the channel statistics. We also observe that the black-box NN requires most channel samples for training while the proposed deep-unfolding NN needs much fewer training samples since it fully exploits the structure of our proposed SSCA-based mixed-timescale beamforming algorithm\;\!\footnote{Note that the number of training samples required for the deep-unfolding NN and the black-box NN are suitable for the offline training stage. For the case of online deployment, since the channel statistics vary continuously between adjacent time blocks, transfer learning and meta learning can be used to significantly reduce the required training samples and time in each coherence time block.}.   

Table~\ref{table:layer} presents the weighted sum-rate performance versus the number of layers. Similarly, the results are all normalized by a reference value, which is the weighted sum-rate of the SSCA-based algorithm with 100 layers. Note that the number of layers of the SSCA-based algorithm and the black-box NN refer to the maximum iteration number of the BCD-type short-term active beamforming algorithm $I_{max}$ and the number of FCLs (each layer with $1000$ neurons), respectively. We observe that the performance of the SSCA-based algorithm monotonically increases with $I_{max}$ and when $I_{max}=10$, it achieves $90.21\%$ performance of when it converges. We can also  see that the performance of the deep-unfolding NN increases with the number of layers $I_{u}$ when it is small. When $I_{u}$ is greater than $7$, the result fluctuates. Hence, $I_{u}$ could be selected as $7$ or $8$ since it achieves a good balance between the performance and computational complexity. For the black-box NN, increasing the number of layers  does not significantly improve the performance.
\begin{table*}[t]
	\centering
	\caption{Weighted sum-rate performance versus different numbers of AP antennas ($N$).}
	\begin{tabular}{c|cccccc}
		  \hline
		$N$&8&16&32&64&128&256 \\ 
		  \hline
		SSCA (bits/s/Hz)&$26.49$&$30.70$&$34.73$&$39.19$&$43.89$&$48.65$ \\
		\hline
		Deep-unfolding NN&$98.94\%$&$98.32\%$&$97.71\%$&$96.97\%$&$96.13\%$&$95.08\%$ \\
		\hline
		Black-box NN&$90.95\%$&$88.65\%$&$86.01\%$&$83.19\%$&$80.42\%$&$77.94\%$ \\
		\hline 
	\end{tabular}
	\label{table:antenna} \vspace{0.1em}
\end{table*}
\begin{table*}[htbp] 
	\centering 
	\caption{The CPU running time of the analyzed schemes.}
	\begin{tabular}{c|cc|cccc}
		\hline
		\multirow{2}*{(N,T)} &\multicolumn{2}{c|}{CPU training time (min)}& \multicolumn{4}{c}{CPU testing time (s)} \\
		~ & deep-unfolding & black-box&SSCA & deep-unfolding & black-box&full CSI  \\
		\hline
		(8,100) &10.80&35.77&2.82&0.052&0.016&18.62 \\ \hline
		(16,100) &11.53&38.85&2.90&0.054&0.017&18.71 \\ \hline
		(32,100) &15.89&41.87&3.07&0.056&0.019&18.89 \\ \hline
		(64,100) &35.75&57.00&3.65&0.058&0.021&19.47 \\ \hline
		(128,100) &52.37&126.54&6.21&0.071&0.027&22.03 \\ \hline
		(256,100) &102.15&267.31&21.45&0.15&0.071&37.33 \\ \hline
		(256,200) &119.14&289.63&21.45&0.15&0.071&66.26 \\ \hline
		(256,300) &137.46&312.34&21.45&0.15&0.071&250.58 \\ \hline
		(256,400) &159.23&343.77&21.45&0.15&0.071&1085.18 \\ \hline
	\end{tabular}
	\label{table:time} 
\end{table*}

Table~\ref{table:antenna} shows the weighted sum-rate performance versus the number of antennas at the AP ($N$). The sum-rate performance of the deep-unfolding NN and the black-box NN is normalized by the corresponding sum-rate of the SSCA-based algorithm. When $N$ is small, the deep-unfolding NN achieves very close performance compared to the SSCA-based algorithm. It suffers from a slight performance degradation with the increase of $N$. However, it still achieves more than $95\%$ performance of the SSCA-based algorithm and is significantly better than the black-box NN. 

Table~\ref{table:time} compares the CPU training time and the testing time of different schemes when the number of antennas at the AP ($N$) and the number of reflecting elements at the IRS ($T$) change.  We observe that the training time of the deep-unfolding NN is less than that of the black-box NN since it fully exploits the structure of the proposed SSCA-based mixed-timescale beamforming design algorithm.  
In terms of the testing time, the proposed deep-unfolding NN and the black-box NN provide a significant advantage over the SSCA-based algorithm and the full CSI scheme, which demonstrates the efficiency of the learning-based approaches.
Moreover, the testing time of the full CSI scheme increases dramatically with $T$ while the testing time of the other mixed-timescale algorithms remains the same. This validates that the mixed-timescale beamforming scheme is much more suitable for practical design.
  
\begin{figure}[!t]
	\centering
	\scalebox{0.63}{\includegraphics{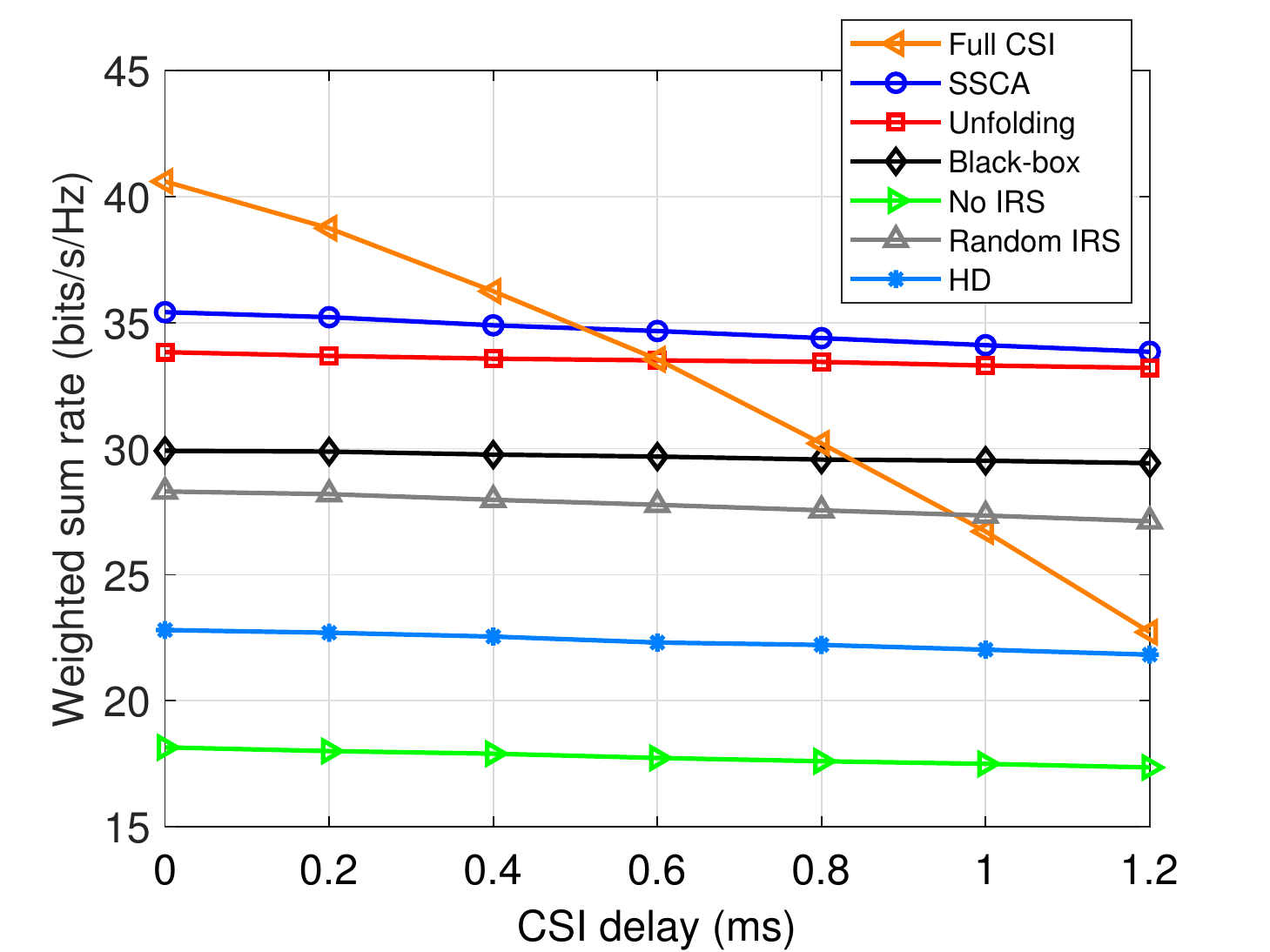}}
	\caption{The weighted sum-rate performance versus the CSI delay $\tau$. }\label{fig:performance_csidelay} \vspace{-0.3em}
\end{figure}

Then, we investigate the impact of the CSI delay $\tau$ on different schemes. We adopt the delay model in \cite{CSI_delay_model} and assume that the CSI delay is proportional to the number of CSI signaling bits~\cite{FDrelay2}.  If the  CSI delay of the single-timescale algorithm is given by $\tau$, then, that of out proposed mixed-timescale algorithm can be computed as $\tau_{m} = \frac{Q_m}{Q_s}\tau$. Fig.~\ref{fig:performance_csidelay} shows the weighted sum-rate performance of different schemes versus the CSI delay $\tau$. As we can see, the proposed mixed-timescale algorithms are insensitive to the CSI delay while the single-timescale scheme relied on the full CSI suffers from severe performance degradation. When the CSI delay is greater than $0.6$ ms, the proposed SSCA-based mixed-timescale beamforming algorithm and deep-unfolding NN outperform the single-timescale algorithm. Moreover, compared with the conventional optimization based algorithm, the deep-unfolding NN and black-box NN are even more robust to the CSI delay. This is because the NN-based algorithms can learn CSI errors from the data and alleviate the performance deterioration. 
\begin{figure}[!t] 
	\centering 
	\scalebox{0.63}{\includegraphics{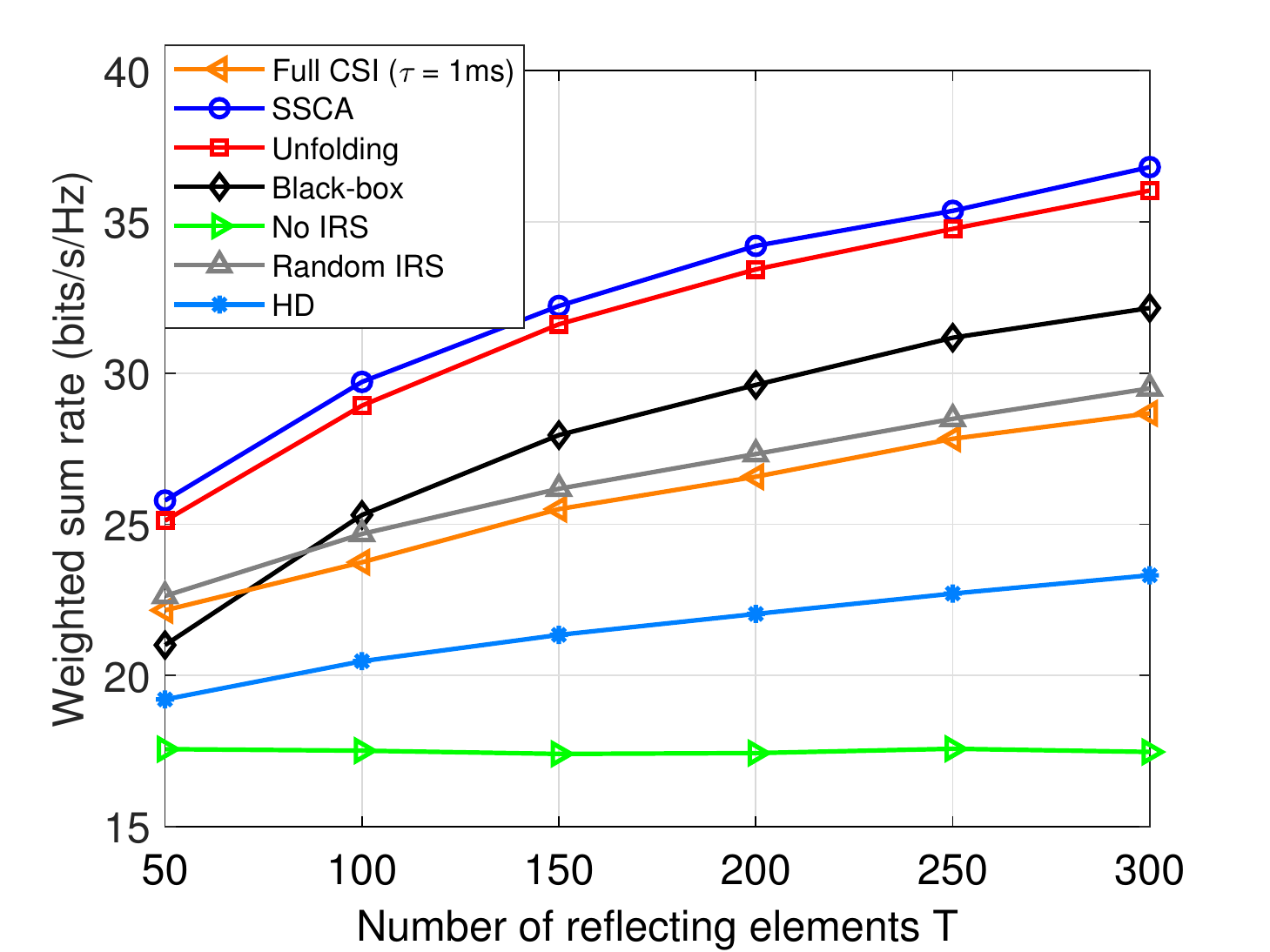}}
	\caption{The weighted sum-rate performance versus the number of reflecting elements $T$. }\label{fig:performance_element} \vspace{-0.3em}
\end{figure}
\begin{figure}[!t] 
	\centering
	\scalebox{0.63}{\includegraphics{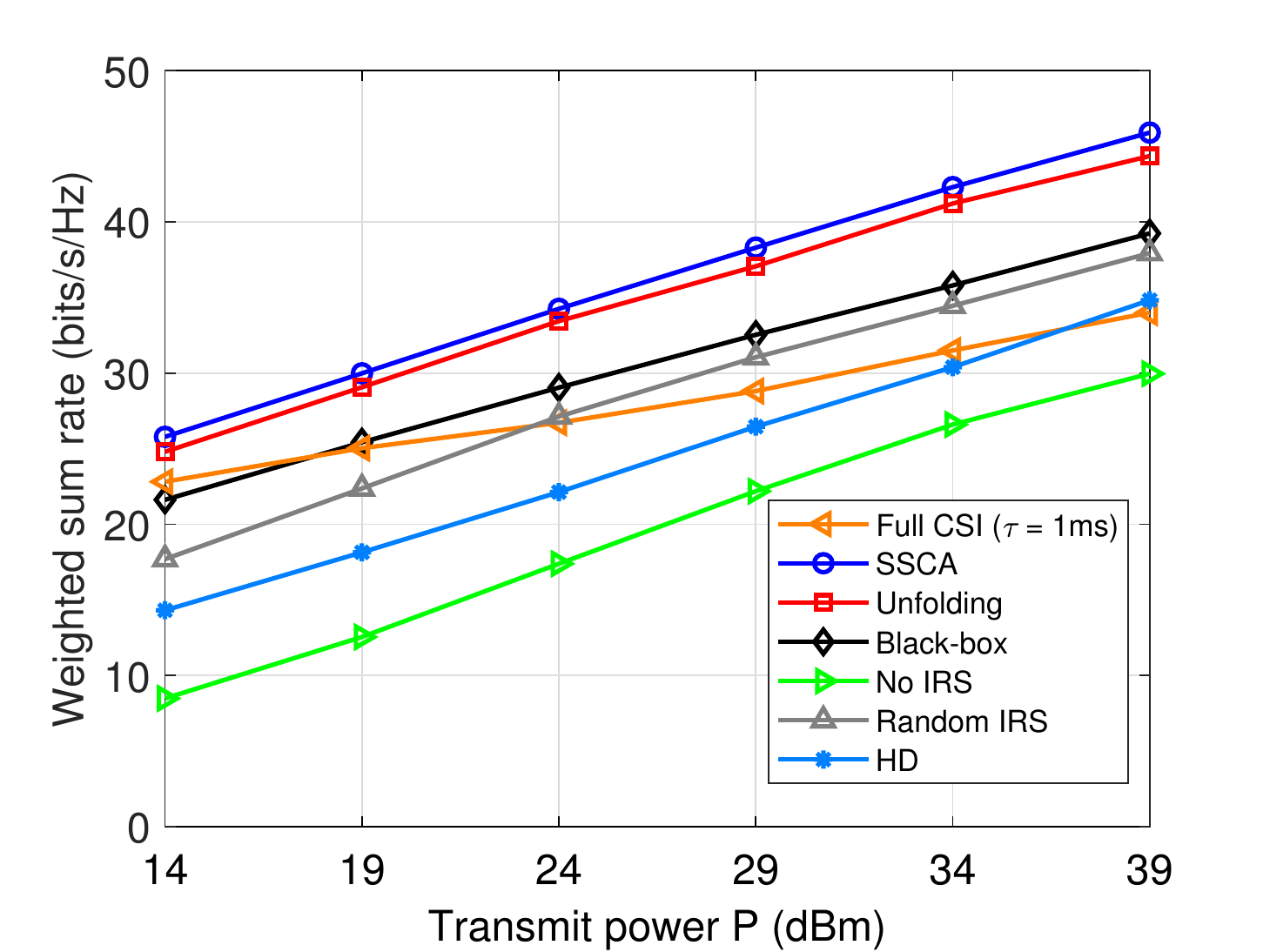}}
	\caption{The weighted sum-rate performance versus different transmit power $P$.}\label{fig:performance_power} \vspace{-0.3em}
\end{figure}

Fig.~\ref{fig:performance_element} presents the weighted sum-rate of different schemes versus the number of reflecting elements of the IRS. As we can see, the proposed deep-unfolding NN approaches the SSCA-based optimization algorithm and it significantly outperforms the other schemes. It is also observed that the weighted sum-rate performance achieved by the SSCA-based algorithm, the deep-unfolding NN, and the black-box NN increases more rapidly with $T$ compared with the random IRS scheme. This is due to the fact that the joint active and passive beamforming design provides a remarkable gain. The full CSI scheme suffers from severe performance degradation due to CSI mismatches and is only comparable with the random IRS scheme. Moreover, the FD scheme provides a huge gap over the HD scheme and the gap increases with $T$. Furthermore, the scheme without IRS provides the worst performance among all the analyzed algorithms, which demonstrates that the IRS can tremendously enhance the spectral efficiency of the conventional FD systems. 

	
%

Fig.~\ref{fig:performance_power} illustrates the performance under different transmit power $P$. Note that we set the power budget of the UL users as $P_{\text{U},k} = P,\forall k$ and that of the AP as $P_{AP} = P+20$ dB. As we can see, the performance of different schemes increases almost linearly with the transmit power (except the full CSI scheme since the CSI error deteriorates its performance). We also observe that the proposed SSCA-based algorithm and deep-unfolding NN both achieve better performance compared with the other schemes under different values of transmit power, which validates the effectiveness of our proposed design.


\begin{figure}[!t]
	\centering
	\scalebox{0.63}{\includegraphics{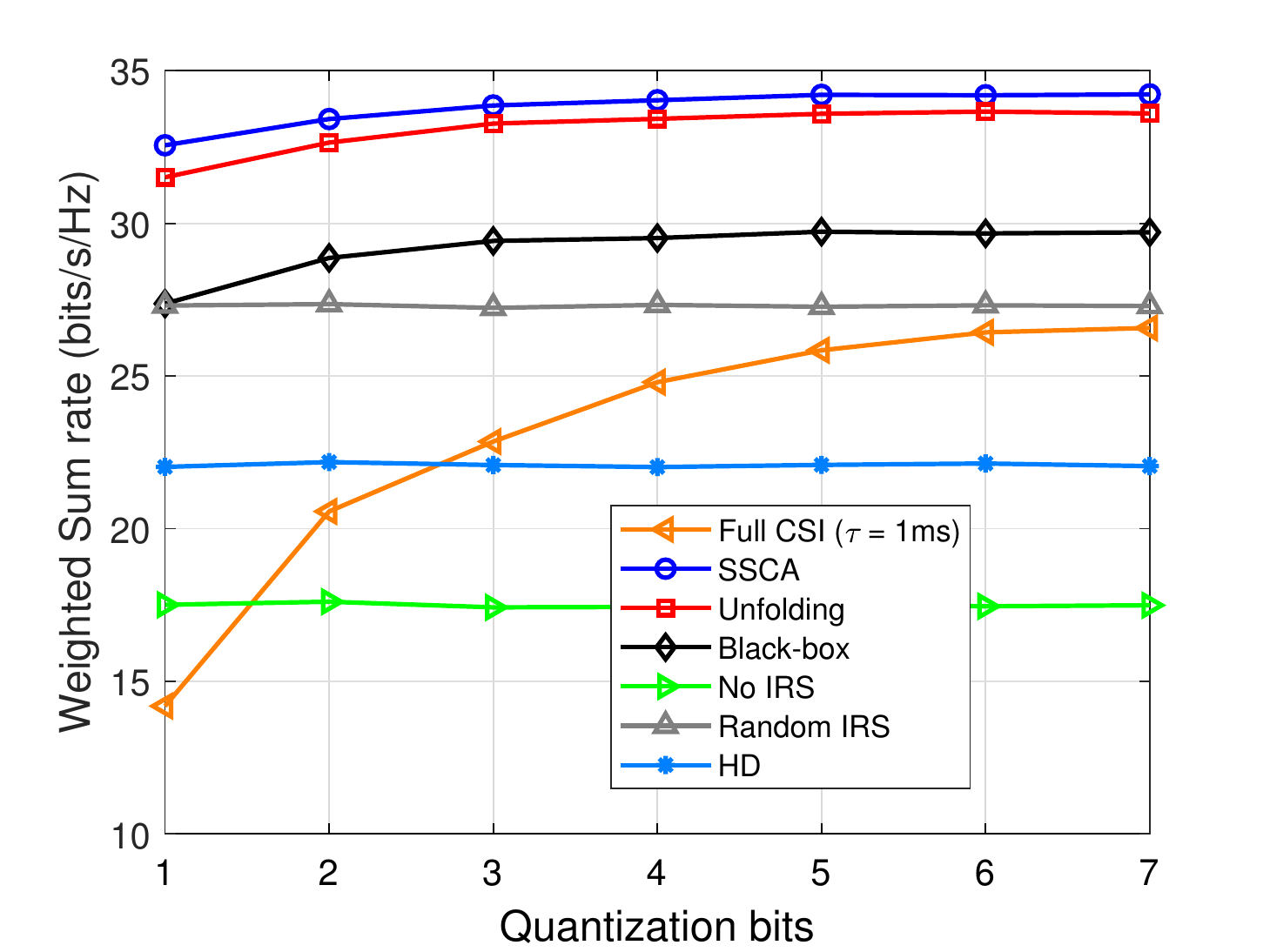}}
	\caption{The weighted sum-rate performance versus different numbers of quantization bits.}\label{fig:quantization} \vspace{-0.0em}
\end{figure}

Fig.~\ref{fig:quantization} shows the sum-rate performance versus different numbers of quantization bits of the IRS. From the figure, the proposed SSCA-based algorithm, the proposed deep-unfolding NN, and the black-box NN are not sensitive to the quantization bits and they can achieve near optimal performance with only a few quantization bits. The quantization of the IRS phase shifters has little effect on the  HD scheme, the random IRS scheme because the phase shifters of the IRS are random. It is also observed that the full CSI scheme is most sensitive to the quantization bits. The sum-rate performance is very poor when there are few quantization bits. This is because in the single-timescale scheme,  $\boldsymbol{\theta}$, $\mathbf{P}_{k}$, and $\mathbf{F}_{l}$ are optimized alternatively based on the full CSI. Thus, the optimality of $\mathbf{P}_{k}$ and $\mathbf{F}_{l}$ relies on the $\boldsymbol{\theta}$ with infinite precision. When $\boldsymbol{\theta}$ is quantized, the derived $\mathbf{P}_{k}$ and $\mathbf{F}_{l}$ are not optimal. In comparison, in the mixed-timescale scheme, $\mathbf{P}_{k}$ and $\mathbf{F}_{l}$ are optimized based on the effective CSI which consists of the full CSI and the quantized $\boldsymbol{\theta}$. The optimality of $\mathbf{P}_{k}$ and $\mathbf{F}_{l}$ holds for the quantized $\boldsymbol{\theta}$. Thus, the proposed mixed-timescale scheme is more robust to the quantization error of the IRS phase shifters.

\begin{figure}[!t]
	\centering
	\scalebox{0.63}{\includegraphics{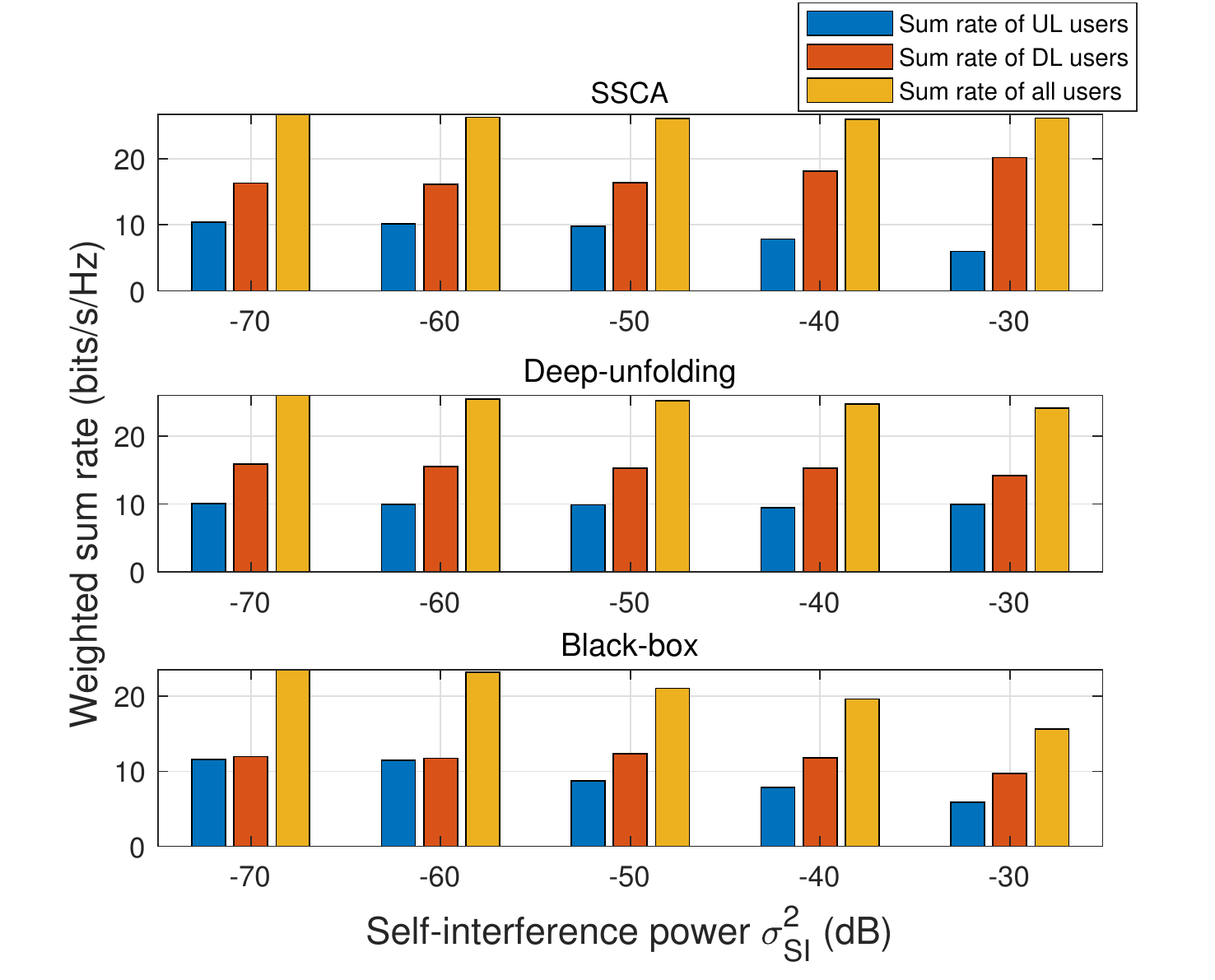}}
	\caption{The weighted sum-rate performance versus the self-interference power $\sigma_{SI}^2$ ($N = 8$).}\label{fig:SI} 
\end{figure}

Fig.~\ref{fig:SI} illustrates the sum-rate performance under different levels of self-interference power when $N=8$. From the figure, the overall sum-rate performance of the SSCA-based algorithm does not change much when the self-interference increases. However, when $\sigma_{SI}^2$ is larger than $-40$ dB, the gap between the UL and DL users becomes larger. When the self-interference increases, the overall sum-rate performance of the proposed deep-unfolding NN decreases slightly, but the deep-unfolding NN still achieves a relatively balanced sum-rate performance between the UL and DL users. Therefore, the proposed deep-unfolding NN can provide a better quality of service (QoS) for the UL users, especially when the self-interference is strong. As for the black-box NN, the overall sum-rate performance suffers from a severe performance deterioration when $\sigma_{SI}^2$ becomes larger. 

\begin{table*}[htbp]
	\setlength\tabcolsep{2.8pt}
	\centering
	\caption{Weighted sum-rate performance versus random locations of users.}
	\begin{tabular}{c|ccccccccc}
		\hline  
		$R_0\, \text{(m)}$&0&2&4&6&8&10 \\ 
		\hline  
		SSCA&$34.73\, (100\%)$&$34.45\,(99.16\%)$&$34.14\,(98.26\%)$&$33.92\,(97.62\%)$&$33.73\,(97.09\%)$&$33.64\,(96.81\%)$ \\
		\hline
		Deep-unfolding NN&$33.94\,(100\%)$&$33.81\,(99.61\%)$&$33.68\,(99.23\%)$&$33.57\,(98.89\%)$&$33.50\,(98.69\%)$&$33.44\,(98.51\%)$& \\
		\hline
		Black-box NN&$29.87\,(100\%)$&$29.84\,(99.89\%)$&$29.80	\,(99.76\%)$&$29.76\,(99.63\%)$&$29.65\,(99.27\%)$&$29.51\,(98.77	\%)$& \\
		\hline 
	\end{tabular}
	\label{table:location}
\end{table*}

Table~\ref{table:location} shows the weighted sum-rate performance when the locations of the users are random. Specifically, the users are randomly located in a circle centered at its original position with radius $R_0$. Note that the percentages in the brackets denote the sum-rate value normalized by the first column. From the table, the weighted sum-rate of all schemes  decreases with $R_0$. The weighted sum-rate of the SSCA-based algorithm declines the fastest and when $R_0 = 10$ m, it achieves $96.81\%$ of the performance of fixed locations. The deep-unfolding NN and the black-box NN have more learning parameters and can adapt to the randomness of the channels well. When $R_0 = 10$ m, they achieve $98.51\%$ and $98.77\%$ of the performance of fixed locations, respectively.

\begin{figure}[!t]
	\centering
	\scalebox{0.63}{\includegraphics{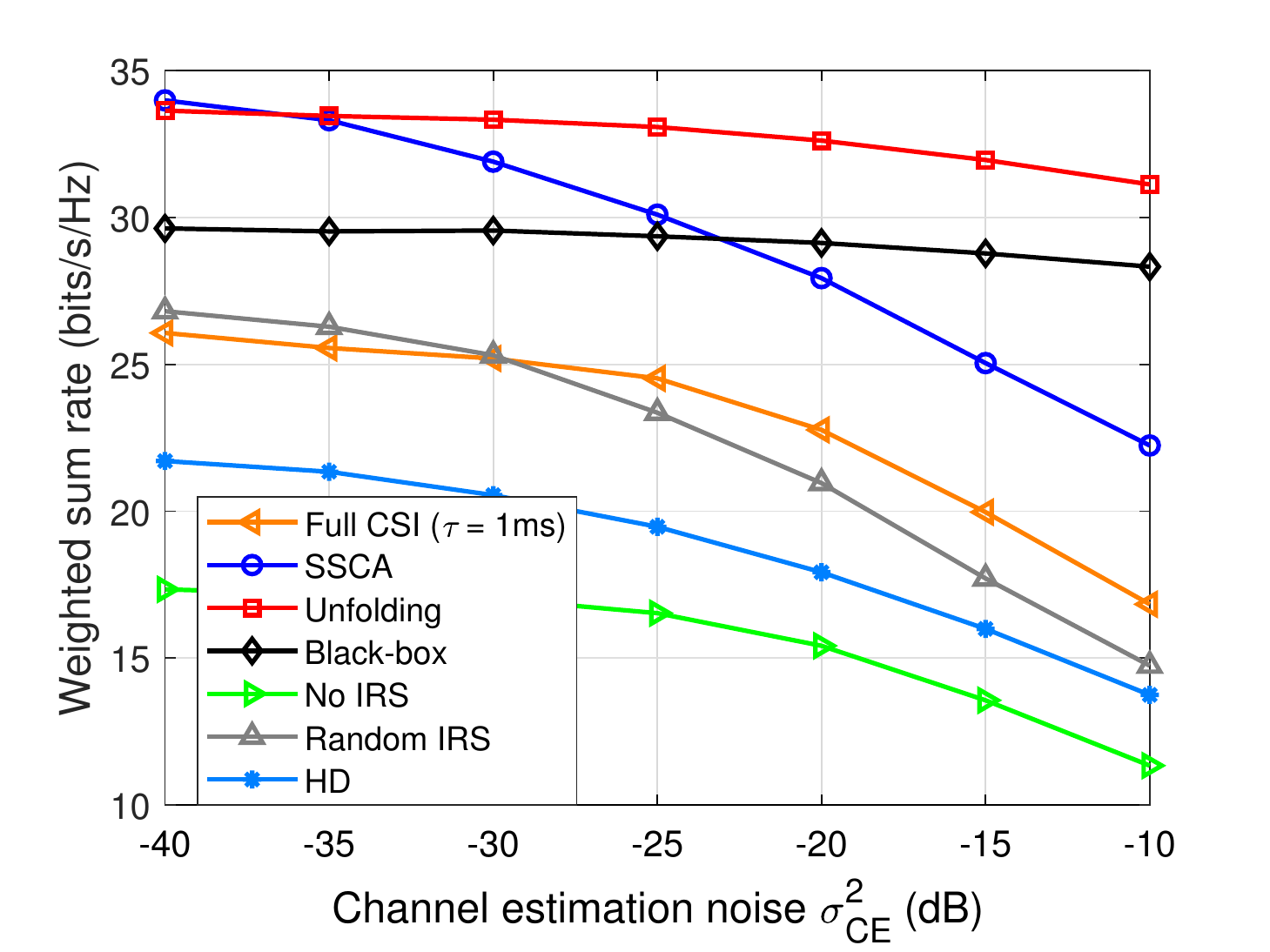}}
	\caption{The weighted sum-rate performance versus  the CSI error variance $\sigma_{CE}^2$.}\label{fig:channel_error} 
\end{figure}

Fig.~\ref{fig:channel_error} presents the achievable weighted sum-rate performance versus the channel estimation error. Specifically, the estimated channel  is model as $\bar{\mathbf{H}} = \mathbf{H}+\bigtriangleup \mathbf{H}$, where $\mathbf{H}$ is the true channel matrix and $\bigtriangleup \mathbf{H}$ is the channel error matrix. We assume that the elements of $\bigtriangleup \mathbf{H}$ 
are independent and follow the Gaussian distribution with zero mean and variance $p_{H}\sigma_{CE}^2$, where $p_{H}$ denotes the average power of the elements in $\mathbf{H}$ and $\sigma_{CE}^2$ indicates the strength of the channel estimation error. From Fig.~\ref{fig:channel_error}, the performance of all schemes degrades with the channel estimation error. Moreover, the weighted sum-rate performance achieved by the conventional optimization based algorithms deteriorates severely as $\sigma_{CE}^2$ increases while the learning based algorithms are much more robust. This is because the learning based algorithms can learn the channel estimation errors from the training samples and alleviate the performance degradation. As we can see, the proposed deep-unfolding NN starts to outperform the SSCA-based algorithm when $\sigma_{CE}^2$ is larger than $-30$ dB, which further demonstrates the benefits of the proposed deep-unfolding design.

\section{Conclusion}
In this paper, we have investigated a MIMO IRS-assisted FD system and formulated a mixed-timescale beamforming design problem for cutting down the heavy CSI overhead. To tackle this highly non-convex optimization problem, an efficient mixed-timescale SSCA-based optimization algorithm has been developed. Moreover, to further reduce the computational complexity of the proposed SSCA-based algorithm, we developed a novel deep-unfolding beamforming algorithm. The deep-unfolding NN consists of a LPBN and a SABN, which maintains the structure of the SSCA-based algorithm but introduces a novel non-linear activation function and some learnable parameters induced by the first-order Taylor expansion to approximate the matrix inversion. It also ties the long-term passive beamforming matrix and the short-term active beamforming matrices more tightly compared with the SSCA-based optimization algorithm. Simulation results verified that the proposed deep-unfolding NN achieves the performance of the SSCA-based optimization algorithm with significantly reduced complexity. 
\vspace{-0.0em}

\begin{appendices}
	\section{Solutions to the Subproblems of the BCD-Type Short-Term Active Beamforming Design} 
	
	\subsubsection{Subproblem w.r.t. $\mathbf{U}_{{\rm U},k},\mathbf{U}_{{\rm D},l}$}
	The subproblems w.r.t. $\mathbf{U}_{{\rm U},k}$ is given by
	\begin{equation}
		\min_{\mathbf{U}_{{\rm U},k}} {\rm Tr} (\mathbf{W}_{\text{U},k}\mathbf{U}_{\text{U},k}^\text{H}\mathbf{A}_{\text{U},k}\mathbf{U}_{\text{U},k}) - 2\Re e \{{\rm Tr}(\mathbf{W}_{\text{U},k}\mathbf{U}_{\text{U},k}^\text{H}\bar{\mathbf{H}}_{\text{U},k}\mathbf{P}_k)\}, \vspace{-0.1em}
	\end{equation}
	where
	\begin{equation}
		\mathbf{A}_{\text{U},k} \triangleq \left(\sum_{k^{'} = 1}^{K}\bar{\mathbf{H}}_{\text{U},k^{'}}\mathbf{P}_{k^{'}}\mathbf{P}_{k^{'}}^{\rm H}\bar{\mathbf{H}}_{\text{U},k^{'}}^{\rm H}+\sum_{l=1}^{L}\tilde{\mathbf{H}}\mathbf{F}_l\mathbf{F}_l^{\rm H}\tilde{\mathbf{H}}^{\rm H}+\sigma_{\text{U}}^2\mathbf{I}\right). \label{AU} \vspace{-0.1em}
	\end{equation}
	By applying the first order optimality condition, the solution of $\mathbf{U}_{{\rm U},k}$ is given by\vspace{-0.1em}
	\begin{equation}
		\mathbf{U}_{{\rm U},k} = \mathbf{A}_{\text{U},k}^{-1} \bar{\mathbf{H}}_{\text{U},k}\mathbf{P}_k. \label{UU_update} \vspace{-0.1em}
	\end{equation}
	Similarly, we obtain the solution of $\mathbf{U}_{{\rm D},l}$ as  \vspace{-0.1em}
	\begin{equation}
		\mathbf{U}_{{\rm D},l} = \mathbf{A}_{\text{D},l}^{-1} \bar{\mathbf{H}}_{\text{D},l}\mathbf{F}_l, \label{UDupdate} \vspace{-0.1em}
	\end{equation}
	where
	\begin{equation}
		\mathbf{A}_{{\rm D},l} \triangleq \left(\sum_{l^{'}=1}^{L}\bar{\mathbf{H}}_{\text{D},l}\mathbf{F}_{l^{'}}\mathbf{F}_{l^{'}}^{\rm H}\bar{\mathbf{H}}_{\text{D},l}^{\rm H}+\sum_{k=1}^{K}\bar{\mathbf{J}}_{k,l}\mathbf{P}_k\mathbf{P}_k^{\rm H}\bar{\mathbf{J}}_{k,l}^{\rm H}+\sigma_{\text{D},l}^2\mathbf{I}\right). \label{AD} \vspace{-0.1em}
	\end{equation}
	\vspace{-0.1em}
	\subsubsection{Subproblem w.r.t. $\mathbf{W}_{\text{U},k},\mathbf{W}_{\text{D},l}$}
	The subproblem w.r.t. $\mathbf{W}_{\text{U},k}$ is given by \vspace{-0.1em}
	\begin{equation}
		\min_{\mathbf{W}_{\text{U},k}} \quad \text{Tr}\left(\mathbf{W}_{\text{U},k}\mathbf{E}_{\text{U},k}\right)-\log \det\left(\mathbf{W}_{\text{U},k}\right). \vspace{-0.1em}
	\end{equation}
	By checking the first order optimality condition, we obtain the optimal solution as \vspace{-0.1em}
	\begin{equation}
		\mathbf{W}_{\text{U},k} =\mathbf{E}_{\text{U},k}^{-1}. \label{WUupdate} \vspace{-0.1em}
	\end{equation}
	Similarly, the optimal solution for $\mathbf{W}_{\text{D},l}$ can be derived as \vspace{-0.1em}
	\begin{equation}
		\mathbf{W}_{\text{D},l} =\mathbf{E}_{\text{D},l}^{-1}. \label{WDupdate} \vspace{-0.1em}
	\end{equation}
	
	\subsubsection{Subproblem w.r.t. $\mathbf{P}_k$}
	After appropriate rearrangement, we can write the subproblem w.r.t. $\mathbf{P}_k$ as
	\begin{subequations}
		\begin{align}
			\min_{\mathbf{P}_k} \quad &{\rm Tr} (\mathbf{P}_k^{\rm H} \mathbf{A}_{\text{P},k} \mathbf{P}_k) - 2\alpha_k\Re e \{\text{Tr}(\mathbf{P}_k^{\rm H} \bar{\mathbf{H}}_{\text{U},k}^{\rm H}\mathbf{U}_{\text{U},k}\mathbf{W}_{\text{U},k})\}\\
			\text{s.t.} \quad & {\rm Tr} (\mathbf{P}_k^{\rm H} \mathbf{P}_k) \leq P_{\text{U},k}, 
		\end{align} 	\label{subproblemP}
	\end{subequations}
	where 
	\begin{equation}
		\begin{split}
		\mathbf{A}_{\text{P},k} &\triangleq \sum_{k^{'}=1}^{K}\alpha_{k^{'}}\bar{\mathbf{H}}_{\text{U},k}^{\rm H}\mathbf{U}_{\text{U},k^{'}}\mathbf{W}_{\text{U},k^{'}}\mathbf{U}_{\text{U},k^{'}}^{\rm H}\bar{\mathbf{H}}_{\text{U},k} \\
		&\qquad+ \sum_{l=1}^{L}\beta_l \bar{\mathbf{J}}_{k,l}^{\rm H}\mathbf{U}_{\text{D},l}\mathbf{W}_{\text{D},l}\mathbf{U}_{\text{D},l}^{\rm H}\bar{\mathbf{J}}_{k,l}. 
     	\end{split}\label{APk} \vspace{-0.2em}
	\end{equation}
	It is readily seen that \eqref{subproblemP} is a convex optimization problem. Therefore, by introducing Lagrange multipliers $\lambda_k \geq 0, \forall k$ and applying the Karush–Kuhn–Tucker (KKT) condition, we can express the optimal solution to $\mathbf{P}_k$ as \vspace{-0.1em}
	\begin{equation}
		\mathbf{P}_k = \alpha_k(\mathbf{A}_{\text{P},k}+\lambda_k\mathbf{I})^{-1} \bar{\mathbf{H}}_{\text{U},k}^{\rm H}\mathbf{U}_{\text{U},k}\mathbf{W}_{\text{U},k}. \label{Pupdate} \vspace{-0.1em}
	\end{equation}   
	Denote  $Q(\lambda_k) = {\rm Tr} (\mathbf{P}_k^{\rm H} \mathbf{P}_k)- P_{\text{U},k}$. If $Q(0) \leq 0$, then we have $\lambda_k = 0$, otherwise, we have $\lambda_k = \lambda_k^*$, where $\lambda_k^*$ is obtained by solving the equation $Q(\lambda_k^*) = 0$ via the bisection search.
	
	\subsubsection{Subproblem w.r.t. $\mathbf{F}_l$}
	Similar to the problem w.r.t. $\mathbf{P}_{k}$, after appropriate rearrangement, we express the subproblem w.r.t. $\mathbf{F}_l$ as \vspace{-0.1em}
	\begin{subequations}
		\label{PBS}
		\begin{align}
			\min_{\mathbf{F}_{l}} \quad & {\rm Tr} (\mathbf{F}_l^{\rm H} \mathbf{A}_{\text{F}} \mathbf{F}_l) - 2\beta_l\Re e \{\text{Tr}(\mathbf{F}_l^{\rm H} \bar{\mathbf{H}}_{\text{D},l}^{\rm H}\mathbf{U}_{\text{D},l}\mathbf{W}_{\text{D},l})\} \\
			\text{s.t.} \quad & {\rm Tr} (\mathbf{F}_l^{\rm H} \mathbf{F}_l) \leq P_{AP},  \vspace{-0.1em}
		\end{align} 
	\end{subequations} 
	where \vspace{-0.1em}
	\begin{equation}
		\begin{split}
		\mathbf{A}_{\text{F}} &\triangleq \sum_{l^{}=1}^{L}\beta_{l^{}}\bar{\mathbf{H}}_{\text{D},l^{}}^{\rm H}\mathbf{U}_{\text{D},l^{}}\mathbf{W}_{\text{D},l^{}}\mathbf{U}_{\text{D},l^{}}^{\rm H}\bar{\mathbf{H}}_{\text{D},l^{}} \\
		&\qquad +\sum_{k=1}^{K}\alpha_k \tilde{\mathbf{H}}^{\rm H}\mathbf{U}_{\text{U},k}\mathbf{W}_{\text{U},k}\mathbf{U}_{\text{U},k}^{\rm H}\tilde{\mathbf{H}}. 
		\end{split}
		\label{AF} \vspace{-0.1em}
	\end{equation}
	By introducing a Lagrange multiplier $\mu \geq 0$ to problem (\ref{PBS}) and employing the KKT condition, we obtain the optimal $\mathbf{F}_l$ as \vspace{-0.1em}
	\begin{equation}
		\mathbf{F}_l = \beta_l(\mathbf{A}_{\text{F}}+\mu \mathbf{I})^{-1}\bar{\mathbf{H}}_{\text{D},l}^{\rm H}\mathbf{U}_{\text{D},l}\mathbf{W}_{\text{D},l},
		\label{Fupdate} \vspace{-0.1em}
	\end{equation}
	where $\mu$ can be found similarly via the bisection search.
\end{appendices}

\end{document}